\documentclass{ptephy_v1}%%%%%% to generate preprint number with ptep logo

\preprintnumber{KEK-TH-2608} %%% %%% Insert preprint number here
\usepackage{amsmath,amssymb,bm}

\numberwithin{equation}{section}

\newcommand{\be}{\begin{equation}}
\newcommand{\ee}{\end{equation}}
\newcommand{\bea}{\begin{eqnarray}}
\newcommand{\eea}{\end{eqnarray}}

\newcommand{\eq}[1]{Eq.~(\ref{eq:#1})}
\newcommand{\sect}[1]{Sec.~\ref{sec:#1}}
\newcommand{\appen}[1]{Appendix~\ref{sec:#1}}

\newcommand{\del}{\partial}

\newcommand{\bra}{\langle}
\newcommand{\ket}{\rangle}
\newcommand{\calO}{{\cal O}}

\newcommand{\Nfour}{${\cal N}=4$}
\newcommand{\SC}{superconductor}
\newcommand{\SCs}{superconductors}
\newcommand{\HSC}{holographic superconductor}
\newcommand{\HSCs}{holographic superconductors}

\bmdefine{\bmq}{{\bm{q}}}
\bmdefine{\bmk}{{\bm{k}}}
\bmdefine{\bmx}{{\bm{x}}}
\bmdefine{\bmy}{{\bm{y}}}
\bmdefine{\bmr}{{\bm{r}}}
\bmdefine{\bmnabla}{{\bm{\nabla}}}
\bmdefine{\bmA}{ \bm{A} }
\bmdefine{\bmD}{ \bm{D} }

\bmdefine{\bmPhi}{ \bm{\Phi} }
\bmdefine{\bmPsi}{ \bm{\Psi} }
\bmdefine{\bmcalO}{ \bm{\mathcal{O}} }
\bmdefine{\bmrho}{ \bm{\rho} }

\newcommand{\calL}{{\cal L}}

\newcommand{\vecx}{\vec{x}}

\newcommand{\epsmu}{\epsilon_{\mu}}

\newcommand{\bmu}{\bar{\mu}}
\newcommand{\bepsmu}{\bar{\epsilon}_\mu}

\newcommand{\bq}{\bar{q}}
\newcommand{\bpsi}{\bar{\psi}}

\newcommand{\bdel}{\bar{\del}}

\bmdefine{\bmg}{{\bm{g}}}
\bmdefine{\bmR}{{\bm{R}}}

\newcommand{\half}{\frac{1}{2}}

\newcommand{\calA}{{\cal A}}

\newcommand{\calF}{{\cal F}}

\newcommand{\calJ}{{\cal J}}
\newcommand{\tilA}{\tilde{A}}

\newcommand{\tilcalA}{\tilde{\calA}}

\newcommand{\tila}{\tilde{a}}
\newcommand{\tilb}{\tilde{b}}

\newcommand{\tilu}{\tilde{u}}
\newcommand{\tilPsi}{\tilde{\Psi}}

\newcommand{\bA}{\bar{A}}
\newcommand{\bPsi}{\bar{\Psi}}

\newcommand{\bcalA}{\bar{\calA}}
\newcommand{\bcalF}{\bar{\calF}}
\newcommand{\bD}{\bar{D}}

\newcommand{\mass}{\text{M}}
\newcommand{\rhou}{U}
\newcommand{\kappaGL}{\kappa_\text{conventional}}

\newcommand{\bo}{1+4A+4B}

\newcommand{\decrease}{\searrow}
\newcommand{\increase}{\nearrow}

\newcommand{\Sos}{\underline{S}}

\newcommand{\at}{\calA_t}

\newcommand{\Cone}{\delta\psi}
\newcommand{\Be}{B_\text{ex}}
\newcommand{\Bc}{B_c}
\newcommand{\Bup}{B_{c2}}
\newcommand{\B}{B}

\begin{document}

\title{What is the dual Ginzburg-Landau theory for holographic superconductors?}%

\author[1]{Makoto Natsuume}
\affil[1]{KEK Theory Center, Institute of Particle and Nuclear Studies, High Energy Accelerator Research Organization, Tsukuba, Ibaraki, 305-0801, Japan 
\thanks{Also at 
Department of Particle and Nuclear Physics, 
SOKENDAI (The Graduate University for Advanced Studies), 1-1 Oho, 
Tsukuba, Ibaraki, 305-0801, Japan;
Department of Physics Engineering, Mie University, 
Tsu, 514-8507, Japan.} \email{makoto.natsuume@kek.jp}}

%\emailAdd{makoto.natsuume@kek.jp}
%\emailAdd{tokamura@kwansei.ac.jp}

%\arxivnumber{2407.13956}
%\preprint{KEK-TH-2608} 
%\pacs{11.25.Tq}

\begin{abstract}%
Holographic superconductors are holographic duals of superconductors. Macroscopically, a superconductor should be described by the Ginzburg-Landau (GL) theory. There is ample evidence that the holographic superconductors are described by the standard GL theory, but the exact form of the dual GL theory is little known. We identify the dual GL theory for a class of bulk 5-dimensional holographic superconductors, where numerical coefficients are obtained exactly.
\end{abstract}

%\begin{document}

%\begin{flushright}
%        \today
%        {\it Preliminary version}
%\end{flushright}

%\keywords{Holography and condensed matter physics (AdS/CMT), AdS-CFT Correspondence, Black Holes}
\subjectindex{AdS/CFT correspondence, Black holes in string theory}

\maketitle

%\flushbottom

%%%%%%%%%
\section{Introduction and summary}%\label{sec:}
%%%%%%%%%

The AdS/CFT duality or the holographic duality \cite{Maldacena:1997re,Witten:1998qj,Witten:1998zw,Gubser:1998bc} is a useful tool to study the ``real world." 
It has been applied to the quark-gluon plasma, hadron physics, nonequilibrium physics, nonlinear physics, and condensed-matter physics (See, \eg, Refs.~\cite{CasalderreySolana:2011us,Natsuume:2014sfa,Ammon:2015wua,Zaanen:2015oix,Hartnoll:2016apf,Baggioli:2019rrs}). 
% v3.2
Among these, the \HSC\ is one of the most studied systems \cite{Gubser:2008px,Hartnoll:2008vx,Hartnoll:2008kx}. 

 The \HSC\ is the holographic dual of superconductors. On the other hand, from the macroscopic point of view, a \SC\ should be described by the Ginzburg-Landau (GL) theory. Then, one of the most basic questions should be: 
%v11
%\begin{center}
%\fbox{
%\begin{tabular}{l}
%What is the dual GL theory for holographic superconductors?
%\end{tabular}
%}
%\end{center}
 \begin{center}
 ``What is the dual GL theory for holographic superconductors?"
 \end{center}
 However, the answer is little known in the literature.
 
There is ample evidence that the \HSC\ is described by the standard GL theory. For example, the very first paper \cite{Hartnoll:2008vx} pointed out that the condensate takes the value of the mean-field critical exponent. This strongly suggests that the \HSC\ is described by  the $|\psi|^4$ mean-field theories (see, \eg, Ref.~\cite{Maeda:2009wv}). 

Identifying the dual GL theory has been initiated in Ref.~\cite{Herzog:2008he} which studied the GL potential terms numerically. Since then, various works appeared, but they are mostly numerical, and the exact form of the GL theory was little known. This is because a \HSC\ is typically an Einstein-Maxwell-complex scalar system. Such a system is hard to solve in general. One often needs either a numerical computation or an approximation method. %The purpose of this paper is to identify the dual GL theory for a class of holographic superconductors.

However, in the bulk 5-dimensions, there exists a simple analytic solution at the critical point for the scalar field $\Psi$ that saturates the Breitenlohner-Freedman (BF) bound \cite{Herzog:2010vz}, and one is able to compute physical quantities analytically.
% v2
%\footnote{The solution is a special case of a one-parameter family of holographic Lifshitz \SCs\ \cite{Natsuume:2018yrg}. }
We compute various physical quantities in the bulk theory and compare them with the GL theory. In this way, we identify the dual GL theory.

A \HSC\ is parameterized by a dimensionless parameter $\mu/T$, where $\mu$ is the chemical potential and $T$ is the temperature. We fix $T$ and vary $\mu$.%
\footnote{When, $\mu=0$, the system is scale invariant so that there is no phase transition.}
Our results are summarized by the following GL free energy:
%\bwt
\begin{subequations}
%\label{eq:}
\begin{align}
f &= \frac{1}{4}|D_i\psi|^2-\frac{\epsilon_\mu}{2}|\psi|^2+\frac{1+4A+4B}{96}|\psi|^4+\frac{1}{4\mu_m}\calF_{ij}^2-(\psi J^*+\psi^* J)~, \\
D_i &=\del_i-i\calA_i~, \\
\mu_m &= \frac{e^2}{1-e^2\ln(\pi T)}~.
%\label{eq:}
%
\end{align}
\end{subequations}
%\ewt
Our notations are explained below, but note that this takes the form of the standard GL theory. The various coefficients are determined because the holographic duality gives a ``first-principle computation."
Here,
\begin{itemize}
\item $\epsmu:=\mu-\mu_c$ is the deviation of the chemical potential from the critical point $\mu_c=2$.
\item $e$ is the $U(1)$ coupling, and $\mu_m$ is the magnetic permeability due to the magnetization current or the normal current (\sect{london}). The value of $\mu_m$ depends on the boundary condition that one imposes.
\item $A$ and $B$ are the parameters in the bulk theory (\sect{nonminimal}). The standard \HSC\ (``minimal holographic superconductor") corresponds to $A=B=0$. The GL theory for the minimal \HSC\ has been proposed in Refs.~\cite{Natsuume:2018yrg,Natsuume:2022kic}.
\item The $T$-dependence is shown explicitly for the $\ln(\pi T)$ term only (see \appen{dimensions} to restore dimensions). 
\end{itemize}
% v2
%Just like in the GL theory, 
This free energy should be regarded as leading terms in the effective theory expansion. There should be the $O(|\psi|^6)$ term and higher, and numerical coefficients are leading ones.

The plan of this paper is as follows:
\begin{itemize}
\item
We first consider the minimal \HSC\ in \sect{minimal}. The system was analyzed previously \cite{Natsuume:2018yrg,Natsuume:2022kic}, but the earlier analysis is not completely satisfactory (\sect{previous}), so we would like to fill the gap. Also, having one paper that collects all materials would be valuable. 
\item
Having computed all physical quantities in the bulk theory, we discuss the dual GL theory in \sect{dual_GL}.
\item 
The analysis of the vortex lattice is rather involved both in the bulk theory and in the GL theory, so we discuss it in a separate section (\sect{vortex} and \appen{vortex_GL}).%
\footnote{See, \eg, Refs.~\cite{Natsuume:2022kic,Maeda:2009vf,Albash:2009iq,Montull:2009fe,Keranen:2009re,Domenech:2010nf,Dias:2013bwa} for holographic vortices.}
\item
Then, we consider the nonminimal \HSCs\ with bulk parameters $A$ and $B$ in \sect{nonminimal}. 
\end{itemize}

%%%%%%%%%
\section{The minimal holographic superconductor}\label{sec:minimal}
%%%%%%%%%

%%------------------
\subsection{Preliminaries}\label{sec:preliminaries}
%%------------------

We consider the bulk 5-dimensional $s$-wave holographic superconductor:%
\footnote{We use upper-case Latin indices $M, N, \ldots$ for the 5-dimensional bulk spacetime coordinates and use Greek indices $\mu, \nu, \ldots$ for the 4-dimensional boundary coordinates. The boundary coordinates are written as  $x^\mu = (t, x^i) =(t, \vecx)=(t,x,y,z)$. 
}
\begin{subequations}
\label{eq:action5}
\begin{align}
S_\text{bulk} &= \frac{1}{16\pi G_5}\int d^5x \sqrt{-g}(R-2\Lambda)+S_\text{m}~, \\
S_\text{m} &= -\frac{1}{g^2} \int d^5x \sqrt{-g} \biggl\{ \frac{1}{4}F_{MN}^2 + |D_M\Psi|^2+m^2 |\Psi|^2 \biggr\}~,
%\label{eq:}
%
\end{align}
\end{subequations}
where 
\begin{align}
F_{MN} =\del_M A_N -\del_N A_M~, \quad
D_M =\nabla_M-iA_M~, \quad
\Lambda =-\frac{6}{L^2}~.
%\label{eq:}
%
\end{align}
We take the probe limit where the backreaction of the matter fields onto the geometry is ignored:
\begin{align}
\frac{1}{g^2N_c^2} \ll 1~,
%\label{eq:}
%
\end{align}
where $N_c$ is the number of ``colors" $N_c^2=8\pi^2L^3/(16\pi G_5)$.
%namely the probe limit corresponds to  the large-$N_c$ limit.
In the probe limit, the matter fields decouple from gravity, and the background metric is given by the Schwarzschild-AdS$_5$ (SAdS$_5$) black hole:
\begin{subequations}
%\label{eq:}
\begin{align}
ds_5^2 &= r^2(-fdt^2+dx^2+dy^2+dz^2)+\frac{dr^2}{r^2f} \\
&=\frac{r_0^2}{u}(-fdt^2+dx^2+dy^2+dz^2)+\frac{du^2}{4u^2f}~, \\
f &= 1-\left(\frac{r_0}{r}\right)^4=1-u^2~, 
%\label{eq:}
%
\end{align}
\end{subequations}
where $u:=r_0^2/r^2$.
For simplicity, we set the AdS radius $L=1$ and the horizon radius $r_0=1$. The Hawking temperature is given by $\pi T =r_0/L^2$. 
The bulk matter equations are given by
\begin{subequations}
%\label{eq:}
\begin{align}
0 &= D^2\Psi-m^2\Psi~, \\
0 &= \nabla_NF^{MN} - J^M~,\\
J_M &= -i\{ \Psi^* D_M\Psi -\Psi(D_M\Psi)^*\} 
= 2\Im(\Psi^* D_M\Psi)~.
%\label{eq:}
%
\end{align}
\end{subequations}

In the $A_u=0$ gauge, the $u\to 0$ asymptotic behaviors of matter fields are given by
\begin{subequations}
\label{eq:dictionary}
\begin{align}
A_\mu &\sim \calA_\mu + A_\mu^{(+)} u~, \\
\Psi &\sim \Psi^{(-)} u^{\Delta_-/2} + \Psi^{(+)} u^{\Delta_+/2}~, \\
\Delta_\pm &:=2\pm\sqrt{4+m^2}~.
%\label{eq:}
%
\end{align}
\end{subequations}
$\calA_t=\mu$ is the chemical potential, and $A_t^{(+)}$ represents the charge density $\bra\rho\ket$. 
Similarly, $\calA_i$ is the vector potential, and $A_i^{(+)}$ represents the current density $\bra\calJ_i\ket$. 
$\Psi^{(+)}$ represents the order parameter $\bra\calO\ket$, and $\Psi^{(-)}$ is the external source for the order parameter.

In this paper, we consider the scalar mass that saturates the BF bound \cite{Breitenlohner:1982bm}:
\begin{align}
m_\text{BF}^2=-4~,
%\label{eq:}
%
\end{align}
or the scaling dimension $\Delta_+=2$.
Then, the asymptotic behavior of $\Psi$ is replaced by
\begin{align}
\Psi \sim \frac{J}{2} u\ln u + \Psi^{(+)} u~.
\label{eq:dictionary1}
\end{align}
According to the standard AdS/CFT dictionary,
\begin{subequations}
\label{eq:dictionary2}
\begin{align}
\bra \calJ^\mu \ket &= \frac{1}{g^2}\left.\sqrt{-g}F^{u\mu} + (\text{counterterm}) \right|_{u=0}~, \\
\psi =\bra \calO \ket &= -\frac{1}{g^2}\,\Psi^{(+)}~,
%\label{eq:}
%
\end{align}
\end{subequations}
where one needs a standard counterterm action for the scalar field and for the Maxwell field. 
%Although we take the probe limit, 
We set the bulk scalar charge $g=1$ below for simplicity.

At high temperature, the equations of motion admit a solution
\begin{align}
A_t = \mu(1-u)~, \quad
A_i = 0~, \quad
\Psi =0~. 
\label{eq:sol_high}
\end{align}
A \HSC\ has 2 dimensionful quantities $T$ and $\mu$, so the system is parameterized by a dimensionless parameter $\mu/T$. We fix $T$ and vary $\mu$. The $\Psi=0$ solution becomes unstable at the critical point and is replaced by a $\Psi\neq0$ solution. For $m^2=-4$, there exists a simple analytic solution at the critical point $\mu_c=2$ \cite{Herzog:2010vz}:
\begin{align}
\Psi \propto -\frac{u}{1+u}~,  \quad\text{at}\quad \mu_c=\Delta=2~.
\label{eq:herzog}
\end{align}
Below we utilize this solution to explore the system.

\paragraph{Counterterms:}
In the bulk 5-dimensions, one needs the counterterm action for the Maxwell field to cancel the UV divergences:
\begin{align}
S_\text{CT}= -\int d^4x\, \frac{1}{4g^2}\sqrt{-\gamma}\gamma^{\mu\nu}\gamma^{\rho\sigma}F_{\mu\rho}F_{\nu\sigma} \times \ln(u^{1/2}/r_0)~,
\label{eq:CT}
\end{align}
where $\gamma_{\mu\nu}$ is the 4-dimensional boundary metric (the 4-dimensional part of the bulk metric). Then, one obtains 
\begin{align}
\bra \calJ^\mu \ket = \left. \frac{2}{g^2}\del_u A_\mu
- \frac{1}{g^2}\del_\nu(\sqrt{-\gamma}F^{\mu\nu}) \times\ln(u^{1/2}/r_0) \right|_{u=0}~.
%\label{eq:}
%
\end{align}
Note the log term takes the form $\ln\tilde{u}$ if one uses $\tilde{u}:=L/r$. We use $u=(r_0/r)^2=(r_0/L)^2\tilde{u}$, so $\ln\tilde{u}=\ln(u^{1/2}L/r_0)$. For example, for the vector perturbation $A_y\propto e^{iqx}$,
\begin{align}
\bra \calJ^y \ket = \left. \frac{2}{g^2}\del_u A_y
- \frac{1}{g^2}q^2\calA_y\left(\half\ln u-\ln r_0 \right) \right|_{u=0}~.
%+ \frac{1}{g^2}(\omega^2-q^2)\calA_y\left(\half\ln u-\ln r_0 \right) \right|_{u=0}~.
\label{eq:current_dict}
\end{align}

\paragraph{The holographic semiclassical equation:}
We have the boundary $U(1)$ Maxwell field $\calA_i$, but in most holographic applications, it is not dynamical: one adds it as an external source. This is because one usually imposes the Dirichlet boundary condition on the AdS boundary. As a result, there is no Meissner effect in standard holographic superconductors. 
% v3.2
Since the Maxwell field is not dynamical, one often calls this case the ``holographic superfluid."
% v3

The procedure to promote the Maxwell field to the classical dynamical field has been known \cite{Compere:2008us}. Impose the Maxwell equation as the boundary condition:
\begin{align}
\del_j\calF^{ij} =e^2 \bra \calJ^i \ket~.
\label{eq:semiclassical}
\end{align}
Here, all quantities including the $U(1)$ coupling $e$ are the boundary ones. Namely, we impose a ``mixed" boundary condition. Ref.~\cite{Natsuume:2022kic} shows the holographic Meissner effect analytically using the boundary condition. In other words, we add the following action to the boundary CFT:
\begin{align}
S_\text{bdy} = -\int d^4x\, \frac{1}{4e^2}\calF_{ij}^2~.
%\label{eq:}
%
\end{align}

One may be unfamiliar to such a boundary condition. 
% v3.2 corrected
It may be worthwhile to consider the boundary condition from the boundary microscopic point of view. For example, consider the \Nfour\ SYM:
\begin{itemize}
\item
The pure gravity is dual to the \Nfour\ SYM. One often uses the system to discuss QGP.
\item
The Einstein-Maxwell theory is dual to the \Nfour\ SYM with a $U(1)$ background. But the Maxwell field here is added only as an external source. One would use the system to discuss QGP at a finite chemical potential.
\item
By imposing the holographic semiclassical equation, the Einstein-Maxwell theory is dual to the \Nfour\ SYM with a dynamical Maxwell field. One would now use the system to discuss QGP with photon.
\end{itemize}
However, we do not really have QGP in mind in this paper: instead, we consider holographic superconductors.

In the literature, one often imposes either the Dirichlet or the Neumann boundary conditions. But our boundary condition is more generic, and those boundary conditions are obtained from our boundary condition as follows:
\begin{itemize}
\item
The Dirichlet boundary condition with a fixed $\calA_i$ corresponds to the $e\to0$ limit.
\item
The Neumann boundary condition $\bra \calJ^i \ket=0$ corresponds to the $e\to\infty$ limit. 
\end{itemize}
Because we impose a mixed boundary condition, one can discuss both cases simultaneously.
% v3
(See also, \eg, Ref.~\cite{Ecker:2021cvz} for another application).

%%------------------
\subsection{High-temperature phase}\label{sec:high}
%%------------------

%%------------------
\subsubsection{The order parameter response function}%\label{sec:}
%%------------------

In the high-temperature phase, there does not exist a spontaneous condensate solution, but there exists a solution with the order parameter source. We consider such a solution here. Namely, we consider the response to the order parameter source and obtain the ``order parameter response function." This gives interesting physical quantities such as the correlation length and the thermodynamic susceptibility.

At high temperatures, the background solution is given by \eq{sol_high}.
Consider the linear perturbation from the background $\Psi= 0+\delta\Psi$. 
% v2
We consider the perturbation of the form  $e^{iqx}$. 
When $\Psi=0$, $\delta A_t$ and $\delta A_i$ decouple from the $\delta\Psi$-equation, and it is enough to consider the $\delta\Psi$-equation:
\begin{align}
0=\del_u\left( \frac{f}{u}\del_u \delta\Psi \right) + \left[ \frac{A_t^2}{4u^2 f} -\frac{q^2}{4u^2} + \frac{1}{u^3} \right] \delta\Psi~,
\label{eq:scalar_high}
\end{align}
where $A_t =(2+\epsmu)(1-u)$. In the high-temperature phase, $\epsmu<0$. Set $\epsmu\to l^2 \epsmu, q \to l q$, and expand $\delta\Psi$ as a series in $l$:
\begin{align}
\delta\Psi = F_0+l^2F_2+\cdots~.
%\label{eq:}
%
\end{align}
We impose the boundary conditions (1) regular at the horizon (2) no fast falloff other than $F_0$. Namely, the order parameter $\psi$ comes only from $F_0$.
The leading order solution is \eq{herzog}:
\begin{align}
F_0 &=- \Cone\,\frac{u}{1+u} \sim -\Cone\, u~,
\quad(u\to0)~,
%\label{eq:}
%
\end{align}
so the order parameter is given by $\delta\psi$.
At the next order,
\begin{align}
F_2 = \Cone\,\frac{u}{8(1+u)}\{ (q^2-2\epsmu)\ln u + 4\epsmu\ln(1+u) \} \sim \frac{1}{8}\Cone\,(q^2-2\epsmu) u\ln u~,
%\label{eq:}
%
\end{align}
so the asymptotic form with $l\to1$ is given by 
\begin{align}
\delta\Psi &\sim \frac{1}{8}\Cone\,(q^2-2\epsmu) u\ln u - \Cone\, u+\cdots~.
%\quad(u\to0)~.
%\label{eq:}
%
\end{align}
Then, one obtains the response function $\chi_>$, the correlation length $\xi_>$, and the thermodynamic susceptibility $\chi_>^T$:
\begin{subequations}
\label{eq:response_high}
\begin{align}
J &=  \frac{q^2-2\epsmu}{4}\Cone~, \\
\to \chi_> &=\frac{\del \delta\psi}{\del J} = \frac{4}{q^2-2\epsmu} \propto \frac{1}{q^2+\xi_>^{-2}}~, \\
\xi_>^2  &=-q^{-2} = \frac{1}{-2\epsmu}~,
\label{eq:xi>0}  \\
\chi_>^T &= \left.\frac{\del \delta\psi}{\del J}\right|_{q=0} =\frac{2}{-\epsmu}
:= \frac{A_>}{-\epsmu}~,
\\
A_> &=2~.
%\label{eq:}
%
\end{align}
\end{subequations}

%%------------------
\subsubsection{The upper critical magnetic field $\Bup$}\label{sec:Hc2_bulk}
%%------------------

Under a magnetic field, \SCs\ are classified into Type I and Type II \SCs:
\begin{itemize}
\item For a Type I \SC, the superconducting state is completely broken at the thermodynamic critical magnetic field $\Bc$. Below $\Bc$, the homogeneous condensate is favorable compared with the normal state.
\item
For a Type II \SC, the magnetic field can partly enter the material while keeping the superconducting state even above $\Bc$. The magnetic field enters by forming vortices. The superconducting state is completely broken above the upper critical magnetic field $\Bup$.
\end{itemize}
% v3
 Then, whether a \SC\ is Type I or Type II depends on the value of the GL parameter $\kappa$:
 \begin{align}
\kappa^2 = \frac{1}{2}\left(\frac{\Bup}{\Bc}\right)^2~.
%\label{eq:}
%
\end{align}
When $\kappa^2<1/2$, $\Bup<\Bc$, and the material belongs to Type I \SC. 
% v3.2 corrected
When $\kappa^2>1/2$, $\Bup>\Bc$, and the material belongs to Type II \SC. 
We discuss $\Bup$ in this section,  discuss $\Bc$ and $\kappa$ later. 
%Incidentally, it is more traditional to define $\kappa$ as
%\begin{align}
%
%\kappaGL^2: = \frac{\lambda^2}{-\xi_>^2}~.
%\label{eq:}
%
%\end{align}
%But in the standard GL theory, $\kappaGL^2=\kappa_B^2$ (\sect{dual_GL}).

We consider the solution of the form $\Psi=\Psi(\vecx,u), A_t=A_t(\vecx,u), A_y=A_y(\vecx,u)$.  The static bulk equations are given by
\begin{subequations}
%\label{eq:}
\begin{align}
0 &= \del_u\left( \frac{f}{u}\del_u \Psi \right) + \left[ \frac{A_t^2}{4u^2f} + \frac{1}{4u^2} (\del_i-iA_i)^2 + \frac{1}{u^3} \right] \Psi~, \\
0 &=\del_u^2 A_t - \frac{1}{2u^2f}|\Psi|^2A_t + \frac{1}{4uf}\del_i^2A_t~, \\
0 &=\del_u(f \del_u A_y)+\frac{1}{4u}\del_i^2A_y -\frac{|\Psi|^2}{2u^2}A_y + \frac{1}{2u^2} \Im[\Psi^* \del_y\Psi]~,
%\label{eq:}
%
\end{align}
\end{subequations}
where we take the gauge $A_u=0$ and $\del_iA^i=0$. In this gauge, one can set $\Psi=\Psi^*$.
We apply a magnetic field $\B$ and approach the critical point from the high-temperature phase. The scalar field $\Psi$ should have an inhomogeneous condensate at $\Bup$. 
Near $\Bup$, $\Psi$ remains small, and one can expand matter fields as a series in $\epsilon$:
\begin{subequations}
%\label{eq:}
\begin{align}
\Psi(\vecx,u) &= \epsilon\Psi^{(1)}+\cdots~, \\
A_t(\vecx,u) &= A_t^{(0)}+\epsilon^2 A_t^{(2)}+\cdots~, \\
A_y(\vecx,u) &= A_y^{(0)}+\epsilon^2 A_y^{(2)}+\cdots~.
%\label{eq:}
%
\end{align}
\end{subequations}
At zeroth order,
\begin{align}
A_t^{(0)} = \mu(1-u)~,\quad
A_x^{(0)} = 0~,\quad
A_y^{(0)} = \B x~.
%\label{eq:}
%
\end{align}
At first order, one solves $\Psi^{(1)}$. Using the ansatz $\Psi^{(1)}=\chi(x)\rhou(u)$, the $\Psi^{(1)}$ equation becomes 
\begin{subequations}
%\label{eq:}
\begin{align}
& (-\del_x^2+\B^2x^2) \chi = E \chi~, \\
% v11: corrected
& \del_u\left( \frac{f}{u}\del_u \rhou \right) + \left[ \frac{(A_t^{(0)})^2}{4u^2f} + \frac{1}{u^3} \right] \rhou = \frac{E}{4u^2}\rhou~,
%\label{eq:}
%
\end{align}
\end{subequations}
where $E$ is a separation constant. The regular solution of $\chi$ is given by Hermite function $H_n$ as
\begin{align}
\chi=e^{-z^2/2}H_n(z)~, z:= \sqrt{\B}x~,
%\label{eq:}
%
\end{align}
with the eigenvalue $E=(2n+1)\B$. %Below we set $n=0$.
$\B$ takes the maximum value when $n=0$ which gives $\Bup$.

Then, the $\rhou$-equation becomes
\begin{align}
0 &= \del_u\left( \frac{f}{u}\del_u \rhou \right) + \left[ \frac{(A_t^{(0)})^2}{4u^2f} - \frac{\Bup}{4u^2} + \frac{1}{u^3} \right] \rhou~.
%\label{eq:}
%
\end{align}
To obtain the upper critical magnetic field $\Bup$, we need the source-free solution (spontaneous condensate) for $\rhou$. But the equation is just \eq{scalar_high} with the replacement $\Bup\to q^2$, so \textit{the following relation holds exactly}:
\begin{align}
\Bup = \frac{1}{-\xi_>^2}~.
%\label{eq:}
%
\end{align}
Also, we consider the \HSC\ with scalar mass $m^2=-4$, but \textit{the above relation holds exactly for the minimal \HSC\ with arbitrary mass.} 
%Here, the minimal \HSC\ means that the bulk scalar field has only the mass term as the potential such as \eq{action5}. 
Moreover, the relation also holds for the class of nonminimal \HSCs\ with arbitrary mass (\sect{nonminimal}).

% v3
Of course, this relation is well-known in the standard GL theory, but the bulk analysis gives the\textit{stronger} statement. The standard GL theory is the leading order in the effective theory expansion, so it is unclear if the relation holds beyond the leading order.

If we express $\Bup$ by $\epsmu$,
\begin{align}
\Bup &= 2\epsmu+\cdots~.
\label{eq:Hc2}
\end{align}

%%------------------
\subsection{Low-temperature phase}\label{sec:low}
%%------------------

%%------------------
\subsubsection{The background}\label{sec:background}
%%------------------

The solution \eqref{eq:herzog} is the one only at the critical point, and we first construct the background solution in the low-temperature phase. 
The construction has been discussed in Refs.~\cite{Herzog:2010vz,Natsuume:2018yrg}. 

Consider the solution of the form
\begin{align}
\Psi=\Psi(u)~, \quad A_t=A_t(u)~, \quad A_i=A_u=0~.
%\label{eq:}
%
\end{align}
The field equations are given by
\begin{subequations}
%\label{eq:}
\begin{align}
0 &=\del_u^2 A_t - \frac{1}{2fu^2}|\Psi|^2A_t~, \\
0 &= \del_u\left( \frac{f}{u}\del_u \Psi \right) + \left[ \frac{A_t^2}{4u^2f} + \frac{1}{u^3} \right] \Psi~, \\
0&=\Psi^* \Psi'-\Psi^{*'} \Psi~.
%\label{eq:}
%
\end{align}
\end{subequations}
One can set $\Psi$ to be real.
Near the critical point, the scalar field remains small, and one can expand matter fields. Namely, we construct the low-temperature background perturbatively: 
\begin{subequations}
%\label{eq:}
\begin{align}
A_t(u) &= A_t^{(0)}+\epsilon^2 A_t^{(2)}+\epsilon^4 A_t^{(4)} +\cdots~, \\
\Psi(u) &= \epsilon\Psi^{(1)}+ \epsilon^3\Psi^{(3)} +\cdots~.
%\label{eq:}
%
\end{align}
\end{subequations}
We obtain the background up to $O(\epsilon^4)$. 
At zeroth order, 
\begin{align}
A_t^{(0)}=\mu_c(1-u)~.
%\label{eq:}
%
\end{align}
The first order solution is \eq{herzog}:
\begin{align}
\Psi^{(1)} = -\frac{u}{1+u}~.
% \quad\text{at}\quad \mu_c=\Delta=2~.
%\label{eq:}
%
\end{align}
To proceed to higher orders in $\epsilon$, we impose the following boundary conditions:
\begin{enumerate}
\item $\Psi^{(n)}$: no fast falloff other than $\Psi^{(1)}$. This means that the condensate $\psi$ comes only from $\Psi^{(1)}$. At the horizon, we impose the regularity condition. % ($n\geq3$)
%no fast falloff ($n\geq3$) and no slow falloff. The former means that $\psi$ comes only from $\Psi^{(1)}$. The latter is the condition for a spontaneous condensate. At the horizon, we impose the regularity condition.  
\item $A_t^{(n)}$: $A_t^{(n)}=0$ at the horizon.
\end{enumerate}
Namely, we fix the fast falloff $\psi$, and the chemical potential is corrected:
\begin{subequations}
%\label{eq:}
\begin{align}
\Psi &\sim \frac{J}{2} u\ln u - \epsilon u~, \\
\mu &= \mu_c + \epsilon^2 \mu_2 + \epsilon^4 \mu_4\cdots~.
%\label{eq:}
%
\end{align}
\end{subequations}

At $O(\epsilon^2)$,
\begin{subequations}
%\label{eq:}
\begin{align}
A_t^{(2)} &= \mu_2(1-u) - \frac{u(1-u)}{4(1+u)} \\
&\sim \mu_2 + \frac{1}{4}(-1-4\mu_2) u +\cdots~.
%\quad(u\to0)
%\label{eq:}
%
\end{align}
\end{subequations}
$\mu_2$ is an integration constant, but it is fixed at the next order from the source-free condition of $\Psi^{(3)}$. 
At $O(\epsilon^3)$,
\begin{subequations}
%\label{eq:}
\begin{align}
\Psi^{(3)} &= \frac{u^2}{12(1+u)^2} + \frac{1}{4}\left(\frac{1}{24}-\mu_2\right) \frac{u\ln u}{1+u} + \frac{8\mu_2-1}{16}\frac{u\ln(1+u)}{1+u} \\
%\frac{-2u^2+u(1+u)\ln(1+u)}{24(1+u)^2}~, 
&\sim \frac{1}{4}\left(\frac{1}{24}-\mu_2\right) u\ln u + \cdots~.
%\quad(u\to0)~.
%\label{eq:}
%
\end{align}
\end{subequations}

Up to $O(\epsilon^3)$,
\begin{subequations}
%\label{eq:}
\begin{align}
\Psi &\sim \frac{1}{4}\left( \frac{1}{24}-\mu_2 \right)\epsilon^3 u\ln u - \epsilon u~, \\
\mu &= A_t|_{u=0} = 2+\epsilon^2\mu_2 +\cdots~.
%\label{eq:}
%
\end{align}
\end{subequations}
The source of the order parameter is given by
\begin{align}
J^{(3)} =\half\left( \frac{1}{24}-\mu_2 \right)~.
%\label{eq:}
%
\end{align}
To obtain the spontaneous condensate, set %$\mu_2=1/24$. 
\begin{align}
\mu_2=\frac{1}{24}~.
\label{eq:mu2}
\end{align}
Then, 
\begin{align}
\epsilon_\mu &:=\mu-2 =\frac{1}{24}\epsilon^2 +\cdots~.
%\label{eq:}
%
\end{align}
This fixes the overall constant $\epsilon$ of the condensate as
\begin{align}
\epsilon^2 &=24\epsmu+\cdots~.
\label{eq:condensate}
\end{align}

The higher order expressions are too cumbersome to write here, and we only give the asymptotic forms. At $O(\epsilon^4)$,
\begin{align}
A_t^{(4)} &\sim \mu_4(1-u) + \left\{ \frac{5}{288}+\frac{-1+8\mu_2}{32}\ln 2 \right\} u +\cdots~.
%\label{eq:}
%
\end{align}
Again, $\mu_4$ is an integration constant, but it is fixed at the next order.

%%------------------
\subsubsection{The on-shell free energy}\label{sec:free_bulk}
%%------------------

We evaluate the on-shell free energy for the low-temperature background. 
The construction has been discussed in Refs.~\cite{Herzog:2010vz,Natsuume:2018yrg}.

Substituting the bulk equations of motion into the bulk matter action, one obtains the matter on-shell action:
\begin{align}
\Sos = -\int d^4x\, \calA_t A_t^{(+)} + \int d^5x\, \sqrt{-g} g^{tt}g^{uu} A_t^2 |\Psi|^2~.
%\label{eq:}
%
\end{align}
We evaluate the on-shell free energy for the spontaneous condensate or the solution with $J=0$, so the boundary term for $\Psi$ vanishes.

We evaluate the difference of the on-shell free energy between the $\Psi\neq0$ solution and the $\Psi=0$ solution. 
%It turns out that 
$\delta \Sos=0$ at $O(\epsilon^2)$, so one has to evaluate the difference at $O(\epsilon^4)$. 
%We first evaluate the leading term. Then, one needs to take into account the background solution up to $O(\epsilon^4)$.

For the $\Psi\neq 0$ solution, the on-shell action becomes
\begin{align}
\frac{\Sos_{\Psi\neq0}}{\beta V_3} = 4(1+\mu_2)\epsilon^2+ \epsilon^4\left( 4\mu_4+ \mu_2^2-\frac{\mu_2}{4}+\frac{1}{48}\right) +\cdots,
%\frac{S_{\Psi\neq0}}{\beta V_3} = 4+ \frac{\epsilon^2}{6} + \epsilon^4\left( 4\mu_4+\frac{7}{576}\right) +\cdots,
%\label{eq:}
%
\end{align}
where $V_3$ is the 3-dimensional spatial volume and $\beta$ is the periodicity of $t$, namely the inverse temperature. One would obtain $\mu_4$ from the $O(\epsilon^5)$ computation of $\Psi$, 
%We know $\mu_4$ from the $O(\epsilon^5)$ computation of $\Psi$, 
but its explicit form is not necessary to evaluate the on-shell action difference because the $\mu_4$-dependence is the same 
% v3.2
for both the $\Psi\neq 0$ and the $\Psi= 0$ solutions. 

For the $\Psi= 0$ solution,
\begin{align}
A_t=(2+\epsilon^2\mu_2+\epsilon^4\mu_4+\cdots)(1-u)~.
%\label{eq:}
%
\end{align}
In this case, only the boundary action contributes since $\Psi=0$. The on-shell action becomes
\begin{align}
\frac{\Sos_{\Psi=0}}{\beta V_3} &= \mu^2 = 4(1+\mu_2)\epsilon^2+ \epsilon^4\left( 4\mu_4+ \mu_2^2 \right) +\cdots.
%\frac{S_{\Psi=0}}{\beta V_3} &= \mu^2 =4+ \frac{\epsilon^2}{6} + \epsilon^4 \left( 4\mu_4+\frac{1}{576} \right) +\cdots.
%\label{eq:}
%
\end{align}
Thus, the difference is
\begin{subequations}
%\label{eq:}
\begin{align}
\delta\Sos &= \Sos_{\Psi\neq0}-\Sos_{\Psi=0} \\
&= \frac{1-12\mu_2}{48}\epsilon^4 \times \beta V_3+\cdots \\
%&= \frac{1}{96}\epsilon^4 \times \beta V_3+\cdots \\
&= -\delta f_\psi \times \beta V_3~, \\
\delta f_\psi &= -\frac{1}{96}\epsilon^4 = -6\epsmu^2~.
\label{eq:fOS1}
\end{align}
\end{subequations}
$\delta f_\psi<0$, so the $\Psi\neq0$ solution is favorable. It is proportional to $\epsmu^2=(\mu-\mu_c)^2$, which implies the second-order phase transition. Namely, the free energy and its first derivative is continuous, but the second derivative is discontinuous.

%%------------------
\subsubsection{The critical magnetic field $\Bc$}\label{sec:Hc}
%%------------------

The on-shell free energy with $\B$ is similar. One can obtain the thermodynamic critical magnetic field $\Bc$.
In the superconducting phase, $\Psi\neq0$ and $A_y=0$, and it is enough to use the previous result. 
In the normal phase, $\Psi=0$ and $F_{xy}=\B$ which does not depend on $u$, so
\begin{subequations}
%\label{eq:}
\begin{align}
\Sos &= -\int d^5x\, \frac{1}{4}\sqrt{-g}F_{MN}^2 \\
&=  -\int d^4x\, \calA_t A_t^{(+)} -\int d^5x  \frac{1}{4}\sqrt{-g} g^{ij}g^{kl}F_{ik}F_{jl} \\
&= \int d^4x\,\mu^2 -\int d^5x\, \frac{1}{8u}F_{ij}^2 
\label{eq:Bc_tmp} \\
&= \int d^4x\,\mu^2 +\int d^4x\, \frac{1}{8}F_{ij}^2 \ln u~.
%\label{eq:}
%
\end{align}
\end{subequations}
In \eq{Bc_tmp}, the indices are raised and lowered by $\delta_{ij}$ not $g_{ij}$. We evaluate the difference between $\B\neq0$ and $\B=0$, so the chemical potential does not make a contribution: 
\begin{align}
\delta \Sos = \int d^4x\, \frac{1}{4}F_{ij}^2 \ln u~.
%\label{eq:}
%
\end{align}
To cancel the UV divergence, one must add the counterterm action \eqref{eq:CT}:
\begin{align}
S_\text{CT}= -\int d^4x\, \frac{1}{4}\sqrt{-\gamma}\gamma^{ik}\gamma^{jl}F_{ij}F_{kl} \times \ln(u^{1/2}/r_0)~
= -\int d^4x \, \frac{1}{4} F_{ij}^2 \ln(u^{1/2}/r_0)~,
%\label{eq:}
%
\end{align}
where $\gamma_{ij}$ is the 3-dimensional boundary spatial metric, and the indices are raised and lowered by $\delta_{ij}$ not $\gamma_{ij}$ in the last expression. Then, one gets the finite result:
\begin{align}
\delta\Sos+S_\text{CT} 
= \int d^4x\, \frac{1}{4} F_{ij}^2 \left(\frac{1}{2}\ln u-\frac{1}{2}\ln u+\ln r_0 \right) 
=\int d^4x\, \frac{1}{4} \ln r_0 F_{ij}^2~.
\label{eq:SosH}
\end{align}
%As explained in \sect{london}, the boundary $U(1)$ Maxwell field $\calA_i$ is added as a source and is not dynamical in most applications. To make it dynamical, 
Also, we add the boundary Maxwell action (\sect{preliminaries}):
\begin{align}
S_\text{bdy} = -\int d^4x\, \frac{1}{4e^2}\calF_{ij}^2~.
%\label{eq:}
%
\end{align}
Thus,
\begin{align}
\delta \Sos+S_\text{CT}+S_\text{bdy} &=  -\int d^4x\, \frac{1}{4\mu_m}\calF_{ij}^2~,
\label{eq:fB} \\
\mu_m &= \frac{e^2}{1-e^2\ln r_0}~.
%\label{eq:}
%
\end{align}
Then, the net effect of these contributions is to change the magnetic permeability from the vacuum value $\mu_0=e^2$ to $\mu_m$.
Finally, for the boundary Maxwell field, a boundary term must be added to cancel the surface term:
\begin{align}
S_G = -\int d^4x \frac{1}{4\mu_m}\del_i(\calF^{ij}\calA_j)~.
%\label{eq:}
%
\end{align}
The on-shell value is negative twice of \eq{fB}. Therefore,
\begin{align}
\Sos_{\B} = &= +\int d^4x\, \frac{1}{4\mu_m}\calF_{ij}^2 = \frac{1}{2\mu_m}\B^2 \times \beta V_3
\nonumber \\
&=: -\delta f_\B \times \beta V_3~, 
%\label{eq:}
%
\end{align}
We compare this free energy with the free energy in the superconducting phase obtained in \sect{free_bulk}.

%\cf, GL:
%\begin{align}
%
%G_n = -\frac{1}{2\mu_m}H^2\times \beta V_3
%\label{eq:}
%
%\end{align}

The critical magnetic field $\Bc$ is obtained by the condition that the homogeneous condensate is thermodynamically favorable compared with the normal state. Then, 
\begin{align}
\Bc^2 = 12\mu_m\epsmu^2+\cdots~.
\label{eq:Hc}
\end{align}
When $\B<\Bc$, $\delta f_\Psi < \delta f_B$, and the superconducting phase is favorable.

%%------------------
\subsubsection{The penetration length}\label{sec:london}
%%------------------

We discuss the Meissner effect in this section and in \sect{vortex}. We follow Ref.~\cite{Natsuume:2022kic}. Below the critical temperature, a uniform condensate $\Psi=\Psi(u)$ is a solution, and we apply a small magnetic field there. For simplicity, we consider $A_y=A_y(x,u)$ with $A_y\propto e^{iqx}$.%$$B=F_{xy}=\del_xA_y$.

The bulk Maxwell equation becomes
\begin{align}
0=\del_u(f\del_u A_y) -\left( \frac{q^2}{4u} + \frac{|\Psi|^2}{2u^2} \right) A_y~.
%\label{eq:}
%
\end{align}
We impose the boundary conditions (1) regular at the horizon (2) $A_y|_{u=0}=\calA_y$. For now, it looks like to impose the standard Dirichlet boundary condition, but we discuss the other boundary conditions as well. 
One can rewrite the equation as an integral equation:
\begin{subequations}
%\label{eq:}
\begin{align}
A_y &= \calA_y - \int_0^u \frac{du'}{f(u')} \int_{u'}^1 du''\, V(u'') A_y(u'')~, \\
V &=\frac{q^2}{4u} + \frac{|\Psi|^2}{2u^2}~.
%\label{eq:formal} 
%
\end{align}
\end{subequations}

One can solve the integral equation iteratively. 
At the leading order,
\begin{align}
A_y &= \calA_y 
- \calA_y \int_0^u \frac{du'}{f(u')} \int_{u'}^1 du''\, V(u'') +\cdots~,
%\label{eq:formal} 
%
\end{align}
which gives
\begin{subequations}
\label{eq:leading}
\begin{align}
\left. 2\del_uA_y \right|_{u=0} 
& = -2\calA_y\int_0^1 du\, V +\cdots \\
%& = -2\calA_y\int_0^1 du\, \left( \frac{q^2}{4u}+\frac{|\Psi|^2}{2u^2}\right)  +\cdots \\
& = \frac{1}{2}\calA_y(q^2\ln u-\epsilon^2)+\cdots |_{u=0}~,
\end{align}
\end{subequations}
where we use $f(0)=1$ and the background solution (\sect{background}).
Then, from the AdS/CFT dictionary \eqref{eq:current_dict}, one obtains
\begin{subequations}
\label{eq:current4}
\begin{align}
\bra \calJ^y \ket &= \left. 2\del_uA_y-\frac{1}{2}q^2 \calA_y(\ln u-2\ln r_0) \right|_{u=0} \\
&= \left\{ q^2(\ln r_0) - \frac{1}{2}\epsilon^2 +\cdots \right\} \calA_y  \\
&=: (c_n q^2 -c_s\epsilon^2) \calA_y~.
%\label{eq:}
%
\end{align}
\end{subequations}
Here, the $r_0$-dependence is shown explicitly only for the $\ln r_0$ term. The term $c_s$ represents the supercurrent. The term $c_n$ exists even in the pure Maxwell theory with $\Psi=0$. This term can be interpreted as the magnetization current due to the normal current. 
%The $c_s$ term gives the London equation:
%\begin{align}
%
%J_y &=  -\frac{1}{2} \epsilon^2 \calA_y \sim -12\epsmu \calA_y = -\frac{1}{\lambda^2}\calA_y 
%\label{eq:}
%
%\end{align}

As the boundary condition at the AdS boundary, we impose the holographic semiclassical equation \eqref{eq:semiclassical}:
\begin{subequations}
\label{eq:lambda1}
\begin{align}
%
%\bra \calJ^y \ket &= (q^2 c_n -c_s\epsilon^2) \calA_y~, \\
\del_j \calF^{ij} &=e^2 \bra \calJ^i \ket~, \\
\to q^2 \calA_y &= e^2(c_n q^2-c_s\epsilon^2)\calA_y+e^2\calJ_\text{ext}~, \\
\to \calA_y &= \frac{e^2}{q^2(1-c_ne^2)+e^2 c_s\epsilon^2} 
\propto \frac{1}{q^2+\mu_m c_s\epsilon^2} =: \frac{1}{q^2+1/\lambda^2}~, \\
\lambda^2 &= \frac{1}{\mu_m c_s\epsilon^2} = \frac{2}{\mu_m \epsilon^2} =\frac{1}{12\mu_m\epsmu}~,
\label{eq:lambda0}  \\
\mu_m &= \frac{e^2}{1-c_ne^2}~.
%\label{eq:}
%
\end{align}
\end{subequations}
Then, the net effect of the normal current is to change the magnetic permeability from the vacuum value $\mu_0=e^2$ to $\mu_m$.
For $\mu_m>0$, $e^2\ln r_0<1$.

We impose the semiclassical equation as the boundary condition. In the literature, one often imposes the Dirichlet boundary condition and the Neumann boundary condition:
\begin{itemize}
\item
% v3.2
The Dirichlet boundary condition with fixed $\calA_y$ corresponds to the $e\to0$ limit. In this case, $\calA_y$ is not dynamical, so one expects no Meissner effect. In fact, the magnetic permeability $\mu_m=0$, and the penetration length diverges $\lambda\to\infty$.
\item
The Neumann boundary condition $\bra \calJ^i \ket=0$ corresponds to the $e\to\infty$ limit. In this case, the magnetic permeability $\mu_m$ becomes
\begin{align}
\mu_\infty = \mu_m|_{e\to\infty}=-\frac{1}{c_n}~.
%\label{eq:}
%
\end{align}
For $\mu_\infty>0$, $r_0<1$.
\end{itemize}

A few remarks are in order: 
\begin{itemize}
\item
Under the Neumann boundary condition, the current $\bra\calJ^y\ket=0$, so the semiclassical Maxwell equation is absent. But the \HSC\ has a dynamical Maxwell field even under this boundary condition. This was explained in terms of the $S$-duality \cite{Witten:2003ya}. But the interpretation is valid only for the 4-dimensional bulk theory. 

There is an alternative interpretation. The current is the sum of the normal current and the supercurrent as we saw. One may regard the normal current as the induced kinetic term. Then, the dynamical Maxwell field is possible even under the boundary condition.
\item
Previously, the normal current contribution was interpreted as the renormalization of the $U(1)$ charge $e$ \cite{Domenech:2010nf}. In the vacuum, this is the correct interpretation. However, in a medium or at a finite temperature, the Lorentz invariance is broken so that a single renormalization does not work. Instead, it is natural to introduce $\mu_m$ and the electric permittivity $\varepsilon_e$ as in elementary electrodynamics. The medium changes these values from the vacuum values. In this sense, the procedure is a kind of ``renormalization."
\end{itemize}

%%------------------
\subsubsection{The order parameter response function}\label{sec:response_low}
%%------------------

We take the gauge $A_u=0$ and perturb around the low-temperature background:
\begin{subequations}
%\label{eq:}
\begin{align}
\Psi &= \bmPsi+\delta\Psi~, \\
A_t &= \bmA_t+a_t~, \\
A_x &= 0 +a_x~,
%\label{eq:}
%
\end{align}
\end{subequations}
where boldface letters indicate the background solution obtained in \sect{background}. 
We consider the perturbation of the form $e^{iqx}$. First, consider the $u$-component of the $A_M$ equation:
\begin{align}
0=qu a_x'+\bmPsi'(\delta\Psi^*-\delta\Psi) - \bmPsi(\delta\Psi^{*'}-\delta\Psi')~.
%\label{eq:}
%
\end{align}
The $\delta\Psi$ equation is real, so $\delta\Psi^*=\delta\Psi$. Then, one can set $a_x=0$.
% v3.2
The rest of field equations are given by
\begin{subequations}
%\label{eq:}
\begin{align}
0 &= \del_u^2 a_t - \left[\frac{q^2}{4uf}+\frac{\bmPsi^2}{2u^2 f}\right] a_t -\frac{\bmA_t\bmPsi}{u^2 f} \delta\Psi~,\\
0 &= \del_u\left(\frac{f}{u}\del_u\delta\Psi\right) + \left[ \frac{\bmA_t^2}{4u^2 f}-\frac{q^2}{4u^2}+\frac{1}{u^3} \right]\delta\Psi + \frac{\bmA_t\bmPsi}{2u^2 f}a_t~. 
%\label{eq:}
%
\end{align}
\end{subequations}
Set $\epsilon\to l \epsilon, q \to l q$, %(\omega\to l^2 \omega)
 and expand the fields as a series in $l$:
\begin{subequations}
%\label{eq:}
\begin{align}
a_t &= a_t^{(0)} + l a_t^{(1)} + l^2 a_t^{(2)}+\cdots~,\\
\delta\Psi &= F_0 + l F_1+ l^2F_2 +\cdots~.
%a_x &= (1-u^2)^{-i\omega/4} (a_x^{(0)}+l a_x^{(1)} + l^2 a_x^{(2)}+\cdots)~.
%\label{eq:}
%
\end{align}
\end{subequations}
%\paragraph{The Dirichlet BC:}
We impose the following boundary conditions:
\begin{enumerate}
\item $a_t^{(i)}=0$ at the horizon, no slow falloff except $a_t^{(0)}$, and $a_t|_{u=0}=\delta\at$.  %$a_t^{(i)}=0$ at $u=0$
\item $\delta\Psi$: regular at the horizon and the condensate comes only from $F_0$. 
\end{enumerate}
Below we give the $\delta\at=0$ solution for simplicity. 
The solution at the leading order is given by 
%$F_0=\Cone\,u/(1+u), a_t^{(0)}=a_{t0}(1-u), a_x^{(0)} = a_{x0}$. 
\begin{align}
%
%a_t^{(0)} &=a_{t0}(1-u)~,\\
F_0 &= -\Cone\,\frac{u}{1+u}~,\quad a_t^{(0)} =0~.
%\label{eq:}
%
\end{align}
At $O(l)$ and $O(l^2)$,
\begin{subequations}
%\label{eq:}
\begin{align}
a_t^{(1)}
&= - \Cone\, \epsilon \frac{u(1-u)}{2(1+u)}~, \\
F_1 &=a_t^{(2)} =0~, \\
%&=\frac{1}{4} a_{t0} \epsilon \frac{u}{1+u}\ln\frac{u}{(1+u)^2}~,\\
\frac{F_2}{\Cone}
&=
 \frac{6q^2+\epsilon^2}{48}\frac{u\ln u}{1+u} 
- \epsilon^2 \frac{u\ln (1+u)}{6(1+u)}
+ \epsilon^2 \frac{u^2}{4(1+u)^2}~. 
%&\sim  -\frac{1}{8} (q^2+4\epsmu)u\ln u~.
%\label{eq:}
%
\end{align}
\end{subequations}
The asymptotic form is given by
\begin{subequations}
%\label{eq:}
\begin{align}
a_t &\sim  -\half \Cone\,\epsilon u~, \\
%\mfa_{t0} - \left(\mfa_{t0}+\frac{1}{2}\Cone\,\epsilon\right) u~, \\
%a_x &\sim a_{x0}~, \\
\delta\Psi &\sim  \frac{1}{48}\Cone\,(6q^2+\epsilon^2) u\ln u - \Cone\, u~.
%\label{eq:}
%
\end{align}
\end{subequations}

Then, one obtains the response function $\chi_<$, the correlation length $\xi_<$, and the thermodynamic susceptibility $\chi_<^T$:
\begin{subequations}
\label{eq:chi_low1}
\begin{align}
%
%a_t^{(1)} &\sim -\sqrt{6} \epsmu^{1/2} u~, \\
%\delta\rho &\sim  -\frac{1}{8} (q^2+4\epsmu)u\ln u+u~,\\
J &= \frac{q^2+4\epsmu}{4}\Cone~, \\
\to \chi_< &= \frac{\del \delta\psi}{\del J} = \frac{4}{q^2+4\epsmu} \propto \frac{1}{q^2+\xi_<^{-2}}~, \\
%0&= q^2+\xi_<^{-2}~, \quad 
\xi_<^2  &=-q^{-2} = \frac{1}{4\epsmu}~, \\
\chi_<^T &= \frac{1}{\epsmu}
:= \frac{A_<}{\epsmu}~,\\
A_< &=1~.
%\label{eq:}
%
\end{align}
\end{subequations}
Here, we use $\epsilon^2=24\epsmu+\cdots$.
%The charge density is given by
%\begin{align}
%
%\bra \delta\calJ^t \ket 
%&= - 2a_t' = \Cone\,\epsilon~.
%\label{eq:}
%
%\end{align}

%%------------------
\subsubsection{The GL parameter}\label{sec:kappa}
%%------------------

We define the GL parameter $\kappa_B$ as%
\footnote{
$\kappa_B$ is known as the Maki parameter $\kappa_1$ \cite{maki}.}
\begin{align}
\kappa_B^2 &:= \frac{1}{2}\left(\frac{\Bup}{\Bc}\right)^2 = \frac{1}{6\mu_m}~,
%\label{eq:}
%
\end{align}
where we use \eq{Hc2} and \eq{Hc}. 
However, it is more traditional to define $\kappa$ as
\begin{align}
\kappaGL^2: = \frac{\lambda^2}{-\xi_>^2} 
= \frac{1}{6\mu_m}~,
%\label{eq:}
%
\end{align}
where we use \eq{xi>0} and \eq{lambda0}. 
Note that it is conventional to use $\xi_>$ not $\xi_<$ to define $\kappa$. 
% v3.2
If one were to use $\xi_<$, an appropriate definition would be
\begin{align}
\kappa_<^2 := \frac{\lambda^2}{2\xi_<^2}
= \frac{1}{6\mu_m}~,
%\label{eq:}
%
\end{align}
where we use \eq{chi_low1}. 
Note the factor $1/2$.%
\footnote{For example, Ref.~\cite{Dias:2013bwa} seems to compare $\xi_<$ and $\lambda$ without the factor $1/2$. They report the transition from Type II to Type I \SCs\ by changing the bulk scalar charge $g$. The report itself may be valid, but the classification may differ if one takes into account the factor 2.}
%but they should be mostly Type I according to our definition.
It does not matter which definition one chooses because they give the same result in the standard GL theory (\sect{dual_GL}).

\begin{itemize}
\item
Our GL parameter $\kappa$ depends both on the $U(1)$ coupling $e$ and on $\mu_m$ which is temperature dependent. Thus, whether our system is Type I or Type II depends on the values of $e$ and $T$. Of course, $e$ is fixed in the real world, and $\mu_m$ is almost constant in real materials. For simplicity, set $e=r_0=1$. Then, $\kappa^2=1/6$, which means that the system belongs to a Type I superconductor.
\item
For the nonminimal \HSCs\ in \sect{nonminimal}, $\kappa$ depends on the bulk parameters $A$ and $B$.  The system approaches a more Type II-\SC\ like material by choosing $A$ and $B$ appropriately.
\end{itemize}

%%------------------
\subsection{Bulk analysis: differences from previous works}\label{sec:previous}
%%------------------

Our main emphasis in this paper is nonminimal \HSCs, but the expressions for the systems are a little complicated, so we first start from the minimal \HSC.
In addition, the analytic solution of the minimal \HSC\ and its dual GL theory were analyzed previously \cite{Natsuume:2018yrg,Natsuume:2022kic}, but the earlier analysis is not completely satisfactory because 
\begin{itemize}
\item
Previous analysis typically imposes the Dirichlet boundary condition on the AdS boundary. As a result, the boundary Maxwell field is not dynamical, and there is no Meissner effect. We impose the ``holographic semiclassical equation" to make the boundary Maxwell field dynamical. This makes it possible to discuss the penetration length, the critical magnetic fields, and the GL parameter.
% v3
\item
Several quantities has not been evaluated before:
\begin{enumerate}
\item
The thermodynamic critical magnetic field $B_c$. 
\item
The order parameter response function at low temperature (\sect{response_low}).
\end{enumerate}
We also point out that the relation $B_{c2}=1/(-\xi_>^2)$ holds exactly for the minimal holographic superconductors with arbitrary mass. The GL theory is an effective theory, and the relation holds only at leading order in the effective theory expansion. But the relation holds exactly for \HSCs. 

\item 
The vortex lattice analysis in Ref.~\cite{Natsuume:2022kic} was not complete. We extend the analysis to the third order (\sect{vortex}). This is necessary to evaluate the free energy and to show that the most favorable configuration is the triangular lattice. 

Ref.~\cite{Maeda:2009vf} analyzed the vortex lattice previously, but the reference imposes the Dirichlet boundary condition. We impose the holographic semiclassical equation instead. As a result, our free energy completely agrees with the GL theory one. 
Also, the analysis of Ref.~\cite{Maeda:2009vf} is rather involved, and we simplify the analysis considerably by incorporating the hydrodynamic limit from the beginning (\sect{vortex}).
\end{itemize}

%%%%%%%%%
\section{The dual GL theory}\label{sec:dual_GL}
%%%%%%%%%

We consider the following GL theory:
\begin{subequations}
\label{eq:GL4}
\begin{align}
f &= c_K|D_i\psi|^2 - a |\psi|^2 + \frac{b}{2}|\psi|^4+\cdots+\frac{1}{4\mu_m} \calF_{ij}^2  -(\psi J^*+\psi^* J)~,  \\
D_i &= \del_i - i\calA_i~, \quad
a = a_0\epsmu+\cdots~, \quad
b =b_0+\cdots~, \quad
c_K =c_0+\cdots~.
\end{align}
\end{subequations}
% v3.1 typo
In the standard GL theory, $\mu_m=e^2$. Namely, we generalize the GL theory where the material has the magnetization current.
The equations of motion are given by
\begin{subequations}
%\label{eq:}
\begin{align}
0&=-c_KD^2\psi-a\psi+b\psi|\psi|^2-J~,\\
0&=\del_j \calF^{ij}-\mu_m \calJ^i~,\\
\calJ_i &=-ic_K[\psi^* D_i\psi - \psi (D_i\psi^*)] = 2c_K\Im[\psi^* D_i\psi]~.
%\label{eq:}
%
\end{align}
\end{subequations}
%At the leading order, 
There are 3 unknown coefficients $a_0,b_0,c_0$.
The coefficient $c_0$ is actually redundant because one can always absorb it by a $\psi$ scaling. Thus, there are 2 independent parameters. But it is useful to keep it to compare with the holographic result. The scaling changes the AdS/CFT dictionary such as \eq{dictionary1} and \eq{dictionary2}. Also, we do not know the exact normalization ($c_0$ is only the leading normalization). 
%since the scaling changes, \eg, the value of the condensate $|\psi|^2$. %We take the scaling into account after we obtain the final result (\sect{summary_GL}).

\paragraph{Determining coefficients:}
We determine the parameters of the dual GL theory from (1) the order parameter response function at high temperature, and (2) the spontaneous condensate.

In the high-temperature phase $\epsmu<0$, there is no spontaneous condensate. When there is no Maxwell field, the linearized $\psi$ equation is 
\begin{align}
0=-c_K \del_i^2 \psi-a\psi-J~.
%\label{eq:}
%
\end{align}
In the momentum space where $\psi\propto e^{iqx}$, 
\begin{align}
0=(c_Kq^2-a)\psi-J~.
%\label{eq:}
%
\end{align}
% v3.1 typo
One obtains the response function for $\psi$:
\begin{align}
\chi_> := \frac{\del \psi}{\del J} = \frac{1}{c_Kq^2-a}~,
%\label{eq:}
%
\end{align}
and the thermodynamic susceptibility is 
\begin{align}
%
%\lab
\chi_>^T :=\chi_>|_{q=0}= \frac{1}{-a_0\epsmu}=:\frac{A_>}{-\epsmu}~,
%\label{eq:}
%
\end{align}
where $A_>$ is the critical amplitude. The correlation length is given by
\begin{align}
\xi_>^2=-\frac{c_K}{a}=\frac{c_0}{a_0}\frac{1}{-\epsmu}~.
%\label{eq:}
%
\end{align}
From the holographic result \eqref{eq:response_high},
\begin{align}
\chi_>=\frac{4}{q^2-2\epsmu}~.
%\label{eq:}
%
\end{align}
This fixes
\begin{align}
a_0=\half~, \quad c_0=\frac{1}{4}~, 
%\label{eq:}
%
\end{align}
so
\begin{align}
A_>=2~, \quad \xi_>^2= \frac{1}{-2\epsmu}~.
%\label{eq:}
%
\end{align}

In the low-temperature phase $\epsmu>0$, there is a homogeneous spontaneous condensate:
\begin{align}
|\psi_0|^2 = \epsilon^2 = \frac{a}{b} = \frac{a_0}{b_0}\epsmu~.
%\label{eq:}
%
\end{align}
From the holographic result \eqref{eq:condensate}, $|\psi_0|^2=\epsilon^2=24\epsmu$, which fixes
\begin{align}
b_0=\frac{1}{48}~.
%\label{eq:}
%
\end{align}
Thus, the dual GL theory becomes
\begin{align}
%
%F = \int d^3x \left\{ 
f= \frac{1}{4}|D_i\psi|^2-\frac{\epsilon_\mu}{2}|\psi|^2+\frac{1}{96}|\psi|^4+\frac{1}{4\mu_m}\calF_{ij}^2-(\psi J^*+\psi^* J)~.
% \right\}~.
%\label{eq:}
%
\end{align}
The magnetic permeability is given by
\begin{align}
\mu_m = \frac{e^2}{1-(\ln r_0)e^2}~.
%\label{eq:}
%
\end{align}
One can now determine the rest of physical quantities:
\begin{enumerate}
\item The response function at low temperature: the correlation length $\xi_<$, the thermodynamic susceptibility $\chi_<$, and the critical amplitude $A_<$.
\item The penetration length $\lambda$.
\item The on-shell free energy and the thermodynamic critical magnetic field $\Bc$.
\item The upper critical magnetic field $\Bup$.
\item The GL parameter $\kappa$.
\end{enumerate}
In order to make sure that \HSC is  really described by the GL theory, let us derive these quantities and compare them with  holographic results. 

\paragraph{The response function (low temperature):} 
In the low-temperature phase, expand $\psi$ as $\psi=\epsilon+\delta \psi$. The linearized $\delta\psi$-equation is
\begin{subequations}
%\label{eq:}
\begin{align}
0 &=-c_K\del_i^2\delta\psi -a\delta\psi +3b\epsilon^2\delta\psi-J~, \\
\to 
0 &= (c_Kq^2+ 2a)\delta\psi-J~.
%\label{eq:}
%
\end{align}
\end{subequations}
Then, the response function is given by
\begin{subequations}
%\label{eq:}
\begin{align}
\chi_< &:= \frac{\del\delta\psi}{\del J} = \frac{1}{c_Kq^2+2a}~,\\
\xi_<^2 &= \frac{c_K}{2a} = \frac{1}{4\epsmu}~, \\
\chi_<^T &:=\chi_<|_{q=0}= \frac{1}{2a} = \frac{1}{\epsmu} =: \frac{A_<}{\epsmu}~, \\
A_< &= 1~, 
%\label{eq:}
%
\end{align}
\end{subequations}
which agree with the holographic results \eqref{eq:chi_low1}. 
The ratio of critical amplitudes is
\begin{align}
\frac{A_>}{A_<}=2~.
%\label{eq:}
%
\end{align}
%\paragraph{The critical amplitudes:}
%For the homogeneous case, the equation of motion is
%\begin{align}
%
%0=-a\psi+b\psi|\psi|^2-J~.
%\label{eq:GL_source}
%
%\end{align}
%Differentiating the equation with respect to $J$ gives
%\begin{subequations}
%\label{eq:}
%\begin{align}
%
%0&= (-a+3b|\psi|^2)\frac{\del\psi}{\del J}-1 \\
%\to \chi_<^T &= \left.\frac{\del \psi}{\del J} \right|_{J=0} =\frac{1}{2a_0\epsmu} = \frac{1}{\epsmu} := \frac{A_<}{\epsmu}~,
%\label{eq:}
%
%\end{align}
%\end{subequations}
%and $A_<=1$. 

\paragraph{The penetration length:}
For the homogeneous condensate, the $U(1)$ current is
\begin{align}
\calJ_i=-2c_K|\psi_0|^2\calA_i := -\frac{1}{\mu_m\lambda^2}\calA_i~,
%\label{eq:}
%
\end{align}
so the penetration length is
\begin{align}
\lambda^2=\frac{1}{2c_K\mu_m}\frac{b}{a} =\frac{1}{2c_0\mu_m}\frac{b_0}{a_0\epsmu} = \frac{1}{12\mu_m\epsmu}~,
%\label{eq:}
%
\end{align}
which agrees with the holographic result \eqref{eq:lambda1}. 

\paragraph{The on-shell free energy:}
Consider the on-shell free energy. 
In the superconducting phase, $|\psi_0|^2=\epsilon^2=-a/b$ and $\calA_i=0$ due to the Meissner effect, so the on-shell free energy is given by
\begin{align}
f_\psi  = -\frac{b}{2}\epsilon^4 = -\frac{a^2}{2b} = -6\epsmu^2~,
%\frac{G_s}{V_3} =\frac{F_s}{V_3} = -\frac{b}{2}\epsilon^4 \times V_3 = -\frac{a^2}{2b} \times V_3= -6\epsmu^2 V_3~,
%\label{eq:}
%
\end{align}
This agrees with the holographic result \eqref{eq:fOS1}. 

For the Maxwell field, we would like a free energy under a fixed magnetic field. In this case, a boundary term must be added:
\begin{align}
F_G = F -\frac{1}{\mu_m} \int d^3 x\, \del_i(\calF^{ij}\calA_j)~.
%\label{eq:}
%
\end{align}
This is the Gibbs free energy. The variation of $F$ includes the term
\begin{align}
\delta F = \cdots+ \frac{1}{\mu_m} \int d^3x\, \del_i(\calF^{ij} \delta\calA_j)~,
%\label{eq:}
%
\end{align}
so $F$ is appropriate to fix $\calA_i$ on the boundary. On the other hand, the variation of $F_G$ includes the term
\begin{align}
\delta F_G = \cdots+ \frac{1}{\mu_m} \int d^3x\, \del_i(\delta\calF^{ij} \calA_j)~,
%\label{eq:}
%
\end{align}
so $F_G$ is appropriate to fix $\calF_{ij}$ on the boundary.
In the normal phase, $\psi=0$ and $\calF_{xy}=\B$, so 
\begin{align}
F_G = -\frac{1}{4\mu_m}\int d^3x\, \calF_{ij}^2 = -\frac{1}{2\mu_m}\B^2 \times V_3 =: f_B \times V_3~.
\label{eq:gibbs_n}
\end{align}
where $V_3$ is the 3-dimensional volume. 

The critical magnetic field $\Bc$ is defined by the condition that the homogeneous condensate is thermodynamically favorable compared with the normal state $f_\psi < f_\B$. Then,
\begin{align}
\Bc^2=\frac{a^2}{b}\mu_m=12\mu_m \epsmu^2~,
%\label{eq:}
%
\end{align}
which agrees with the holographic result \eqref{eq:Hc}.

\paragraph{The upper critical magnetic field: }%\label{sec:}
The upper critical magnetic field $\Bup$ is discussed in \appen{vortex_GL}:
\begin{align}
\Bup&=\frac{a}{c_K}=2\epsmu~,
%\label{eq:}
%
\end{align}
which agrees with the holographic result \eqref{eq:Hc2}. 

Note that the following relation holds:
\begin{align}
\Bup &= \frac{1}{-\xi_>^2}~.
%\label{eq:Hc2}
%
\end{align}
We saw this in the bulk analysis, but the bulk analysis gives a \textit{stronger} statement. For the holographic superconductor, the relation is exact and holds to all orders in the perturbative expansion in $\epsmu$. The GL theory only shows that the relation holds approximately at the leading order in $\epsmu$. 

\paragraph{The GL parameter:} Then, the GL parameter is given by
\begin{align}
\kappa_B^2 &:= \frac{1}{2}\left(\frac{\Bup}{\Bc}\right)^2 =\frac{b}{2\mu_m c_K^2} =  \frac{1}{6\mu_m}~.
%\label{eq:}
%
\end{align}
The conventional definition gives the same result:
\begin{align}
\kappaGL^2: = \frac{\lambda^2}{-\xi_>^2} = \frac{1}{6\mu_m}~.
%\label{eq:}
%
\end{align}

%%%%%%%%%
\section{The vortex lattice}\label{sec:vortex}
%%%%%%%%%

So far, we consider a homogeneous condensate $\psi=\epsilon$. In this section, we consider an inhomogeneous condensate. We consider the case where the magnetic field is near the upper critical magnetic field $\Bup$. 

In a Type II \SC, the magnetic field can enter the \SCs\ keeping the superconducting state. The magnetic field enters by forming vortices. 
% v2
%As one increases the magnetic field, the magnetic field begins to penetrate into the \SC, and vortices appear at the lower critical magnetic field $H_{c1}$. 
As one increases the magnetic field further, more and more vortices are created, and the vortices form a lattice which is called the vortex lattice. Eventually, the superconducting state is completely broken at the upper critical magnetic field $\Bup$. 
% v2
Such holographic vortex lattices have been investigated in Refs.~\cite{Maeda:2009vf,Natsuume:2022kic}, and we partly follow these references. In \appen{vortex_GL}, we summarize the analogous GL analysis for the reader's convenience. 
% that
Also, the bulk analysis is rather involved, so we summarize the necessary formulae that one needs to evaluate in \appen{vortex_formula}.

We take the gauge $A_u=0$ and $\del_iA^i=0$. The bulk Maxwell equations are given by
\begin{subequations}
%\label{eq:}
\begin{align}
%
%0 &= \calL_t A_t^{(0)}~, \\
0 &= \calL_t A_t  + \frac{1}{4u^2 f} J_t
= \calL_t A_t  - \frac{1}{2u^2 f}|\Psi|^2 A_t~, \\
0 &= \calL_V A_i + \frac{1}{4u^2} J_i
=\calL_V A_i + \frac{1}{2u^2} \Im[\Psi^* D_i\Psi]~, 
%0 &= \del_u (\vec{\del}\cdot\vec{A}^{(0)})~,
%\label{eq:}
%
\end{align}
\end{subequations}
where 
\begin{subequations}
%\label{eq:}
\begin{align}
\calL_t &= \del_u^2 +\frac{1}{4uf} \del_i^2~, \\ 
\calL_V &= \del_u(f\del_u)+\frac{1}{4u} \del_i^2~.
%0 &= \del_u (\vec{\del}\cdot\vec{A}^{(0)})~,
%\label{eq:}
%
\end{align}
\end{subequations}
Near $\Bup$, the scalar field remains small, and one can expand matter fields as a series in $\epsilon$, where $\epsilon$ is the deviation parameter from the critical point:
\begin{subequations}
%\label{eq:}
\begin{align}
\Psi(\vecx,u) &= \epsilon\Psi^{(1)}+ \epsilon^3\Psi^{(3)}+\cdots~, \\
A_t(\vecx,u) &= A_t^{(0)}+\epsilon^2 A_t^{(2)}+\cdots~, \\
A_i(\vecx,u) &= A_i^{(0)}+\epsilon^2 A_i^{(2)}+\cdots~.
%\label{eq:}
%
\end{align}
\end{subequations}
Up to $O(\epsilon)$, the argument is the same as the one for $\Bup$ (\sect{Hc2_bulk}).
%%------------------
%\subsection{Zeroth order}%\label{sec:}
%%------------------
At zeroth order, $\calL_t A_t =0$ and $\calL_V A_i=0$, so %Eqs.~\eqref{eq:eom_bulk} become
\begin{align}
A_t^{(0)} = \mu(1-u)~, 
A_x^{(0)} = 0~, 
A_y^{(0)} = \B_0x~.
%\label{eq:}
%
\end{align}
%\margin{?}
We apply an external magnetic field $\B$. At $\Bup$, a superconducting state just begins to form so that the magnetic induction $B\simeq \Be$. \textit{However, it is important to distinguish $\B$ and $\Be$}. As one lowers the magnetic field, $\B$ begins to differ from $\Be$ due to the Meissner effect as we see in a moment. 

But this effect does not happen in \HSCs\ under the Dirichlet boundary condition. The Maxwell field is not dynamical under the boundary condition. In order to discuss the issue, we impose the semiclassical equation \eqref{eq:semiclassical} as the boundary condition. 

%%------------------
\subsection{First order}%\label{sec:}
%%------------------

At first order, the bulk scalar equation becomes
\begin{align}
0 =& \biggl[ \del_u \left(\frac{f}{u}\del_u \right) + \frac{(A_t^{(0)})^2}{4u^2 f} 
+ \frac{1}{4u^2} \{ \del_x^2 + (\del_y - i\B_0x)^2 \} +\frac{1}{u^3}  \biggr] \Psi^{(1)}~.
%\label{eq:}
%
\end{align}
Using the ansatz
\begin{align}
\Psi^{(1)} = e^{iqy}\chi_q(x) U(u)~,
%\label{eq:}
%
\end{align}
one obtains
\begin{subequations}
%\label{eq:}
\begin{align}
\del_u \left(\frac{f}{u}\del_u U\right)+\left[\frac{(A_t^{(0)})^2}{4u^2 f}+\frac{1}{u^3} \right] U &= - \frac{E}{4u^2} U~, 
%\label{eq:eom_rho} 
\\
\left\{ -\del_x^2+\B_0^2 \left(x-\frac{q}{B_0}\right)^2 \right\} \chi_q  &= E \chi_q~,
%\label{eq:}
%
\end{align}
\end{subequations}
where $E$ is a separation constant. The regular bounded solution is given by Hermite function $H_n$ as
\begin{align}
\chi_q = e^{-z^2/2}H_n(z)~, \quad
z:= \sqrt{B_0}\left(x-\frac{q}{\B_0}\right)~,
%\label{eq:}
%
\end{align}
with the eigenvalue $E=(2n+1)\B_0$.
$\B_0$ takes the maximum value when $n=0$ which gives $\Bup$, so
%Below we set $n=0$, so 
\begin{align}
\chi_q = \exp\left\{-\frac{B_0}{2}\left(x-\frac{q}{\B_0}\right)^2 \right\}~.
%\label{eq:}
%
\end{align}
What we obtained is the ``droplet solution," where the condensate decays exponentially. But superpositions of the droplet solution give rise to a vortex lattice solution where a single vortex is arranged periodically. See, \eg, Ref.~\cite{Maeda:2009vf}. So, consider the general solution
\begin{subequations}
%\label{eq:}
\begin{align}
\Psi^{(1)} &= U(u)\psi^{(1)}(x,y)~,
\label{eq:varphi_0} \\
\psi^{(1)}(x,y) &= \int_{-\infty}^\infty dq\, C(q) e^{iqy} \chi_q(x)~.
%\label{eq:}
%
\end{align}
\end{subequations}
%Here, $U$ is the solution of \eq{eom_rho} with $E=B_0$. 
One can obtain the vortex lattice solution by choosing $C(q)$ appropriately. 

% v2
The first order solution \eqref{eq:varphi_0} satisfies
\begin{align}
(\del_y - iA_y^{(0)}) \Psi^{(1)} = i(\del_x - iA_x^{(0)}) \Psi^{(1)}~,
%\label{eq:}
%
\end{align}
so
\begin{subequations}
%\label{eq:}
\begin{align}
2\Im \left[(\Psi^{(1)})^* D_x^{(0)}\Psi^{(1)} \right] &= -\del_y|\Psi^{(1)}|^2~,\\
2\Im \left[(\Psi^{(1)})^* D_y^{(0)}\Psi^{(1)} \right] &= \del_x|\Psi^{(1)}|^2~,
%\label{eq:}
%
\end{align}
\end{subequations}
or 
\begin{align}
2\Im \left[(\Psi^{(1)})^* D_i^{(0)}\Psi^{(1)} \right] = -\epsilon_i^{~j} \del_j|\Psi^{(1)}|^2~,
\label{eq:bulk_current}
\end{align}
where $\epsilon_{xy}=1$. 

\paragraph{The upper critical magnetic field:}
$\Bup$ is obtained by solving the $U$-equation. The $U$-equation becomes
\begin{align}
0=\del_u \left(\frac{f}{u}\del_u U\right) + \left[ \frac{(A_t^{(0)})^2}{4u^2 f} -\frac{\B_0}{4u^2}+\frac{1}{u^3} \right] U~. 
%\label{eq:}
%
\end{align}
One can construct the solution perturbatively in $\B_0$ just like the high-temperature phase.
Set $\epsmu \to l^2 \epsmu $, $\B_0 \to l^2 B_0$, and expand the field as a series in $l$:
\begin{subequations}
%\label{eq:}
\begin{align}
\rhou &= F_0+l^2F_2+\cdots~, \\
A_t^{(0)} &= (2+\epsmu)(1-u)~,
%\label{eq:}
%
\end{align}
\end{subequations}
We again impose the regularity condition at the horizon and no condensate condition except $F_0=-u/(1+u)$. 

At $O(l^2)$,
\begin{align}
F_2 &=\frac{u}{8(1+u)}\{ (B_0-2\epsmu )\ln u + 4\epsmu\ln(1+u) \} \sim \frac{1}{8}(B_0-2\epsmu)u\ln u~, 
%\label{eq:}
%
\end{align}
so the source-free condition for the order parameter gives
\begin{align}
\B_0=\Bup \sim 2\epsmu~.
%\label{eq:}
%
\end{align}

%%------------------
\subsection{Second order}%\label{sec:}
%%------------------

%The construction so far has been discussed in Ref.~\cite{Maeda:2009vf}. 
% v2
%Let us proceed to the second order solution. We now solve the $A_i^{(2)}$ equation and obtain the current $\bra\calJ_i\ket$. We then impose the holographic semiclassical equation $\del_j\calF^{ij}=e^2\bra\calJ^i\ket$ and show the Meissner effect.

% v2
The Maxwell equation at second order is given by
\begin{subequations}
%\label{eq:}
\begin{align}
%
%0&= \calL_V A_i^{(2)} - J_i^{(0)} \\
0 &= \calL_V A_i^{(2)} + \frac{1}{4u^2} 2\Im[(\Psi^{(1)})^* D_i\Psi^{(1)}]~, \\
&= \calL_V A_i^{(2)} - \frac{1}{4u^2}\epsilon_i^{~j} \del_j|\Psi^{(1)}|^2~,
%+ \del_i(\vec{\del}\cdot\vec{A}^{(2)})~, \\
%0 &= \del_u(\vec{\del}\cdot\vec{A}^{(2)})~,
\label{eq:gauge}
\end{align}
\end{subequations}
where we use \eq{bulk_current}. 
%From \eq{gauge}, $(\vec{\del}\cdot\vec{A}^{(2)})$ does not depend on $u$. Thus, one can choose $\vec{\del}\cdot\vec{A}^{(2)}=0$ by the gauge transformation which does not depend on $u$ so that one can keep the $A_u=0$ gauge.  
In momentum space, 
\begin{subequations}
%\label{eq:}
\begin{align}
0 &= \calL_V A_i^{(2)} - g_i~,\\
\calL_V &= \del_u(f\del_u)-\frac{q^2}{4u}~, \\
g_i &= i\epsilon_i^{~j} q_j \frac{|\Psi^{(1)}|^2}{4u^2}~.
%g_i &= i\epsilon_i^{~j} q_j \widetilde{|\Psi^{(1)}|^2}~,
%\label{eq:}
%
\end{align}
\end{subequations}
%where ``$~\tilde{~}~$" represents a Fourier transformed quantity.
%Note that $ \widetilde{|\Psi^{(1)}|^2}$ is the Fourier transformation of $|\Psi^{(1)}|^2$ and is not $ |\widetilde{\Psi^{(1)}}|^2$.

% v2
%The second order solution can be constructed exactly, but it can be shown that it is a nonlocal function in the boundary direction \cite{Maeda:2009vf}. 
% v2
%This is because holographic results correspond to all orders in effective theory expansion. The GL theory takes only the first few terms in the expansion. In fact, at short wavelength, the London equation is replaced by a nonlocal expression known as the Pippard equation. In order to show the Meissner effect, it is enough to take the long-wavelength $q\to0$ limit.

Using the bulk Green's function $G_V$, the solution is formally written as
\begin{subequations}
\label{eq:Ay2}
\begin{align}
A_i^{(2)} &= a_i - \int_0^1 du'\, G_V(u,u')g_i(u')~, \\
\calL_V G_V(u,u') &= \delta(u-u')~.
%\label{eq:Ay2}
%
\end{align}
\end{subequations}
The first term $a_i$ is the homogeneous solution:
\begin{align}
0=\left\{ \del_u(f\del_u) -\frac{q^2}{4u} \right\}a_i~.
%(\calL_1^*+q^2f)a_i &= 0~,\\
%\calL_1^* &= -\del_*^2~.
%\label{eq:}
%
\end{align}
We impose the following boundary conditions:
\begin{itemize}
\item
$G_V$: (1) regular at the horizon and (2) $G_V(u=0,u')=0$.
% f\del_u G_V=0$?
%$\del_*G_V|_{u_*\to\infty}=0$. 
\item
$a_i$: (1) regular at the horizon and (2) $a_i=\calA_i^{(2)}$ at $u=0$. 
\end{itemize}
%\margin{TBC}
One can rewrite the equation as an integral equation: 
\begin{subequations}
%\label{eq:}
\begin{align}
a_i &=\calA_i^{(2)} -\int_0^u \frac{du'}{f(u')} \int_{u'}^1 du''\, V(u'')a_i(u'')~, \\
V(u) &=\frac{q^2}{4u}~.
%\label{eq:}
%
\end{align}
\end{subequations}
When $q$ is small, one can solve the integral equation iteratively. At $O(q^2)$,
\begin{subequations}
\label{eq:ai}
\begin{align}
a_i &= \calA_i^{(2)} - \calA_i^{(2)} \int_0^u \frac{du'}{f(u')} \int_{u'}^1 du''\,\frac{q^2}{4u''}+\cdots \\
&= \calA_i^{(2)} \left\{ 1+\frac{q^2}{4} \int_0^u du' \frac{\ln u'}{1-u'^2} +\cdots \right\}~, \\
%V &= \frac{q^2}{4u} - \frac{i\omega}{2} + \frac{\epsilon^2}{2(1+u)^2} + \cdots~, \\
2\del_ua_i|_{u=0} &= \frac{q^2}{2}\calA_i^{(2)} \ln u+\cdots |_{u=0}~.
%\label{eq:}
%
\end{align}
\end{subequations}

The Green's function $G_V$ is obtained from 2 independent homogeneous solutions. At $O(q^0)$, the homogeneous solutions are
\begin{subequations}
%\label{eq:}
\begin{align}
A_b &= \half\ln\left(\frac{1-u}{1+u}\right)~, \\
A_h&= 1~, \\
W &:= A_b\del_uA_h-(\del_uA_b)A_h= \frac{1}{f} =: \frac{A}{f}~.
%\label{eq:}
%
\end{align}
\end{subequations}
The solution $A_b$ satisfies the boundary condition at the AdS boundary and $A_h$ satisfies the boundary condition at the horizon. 
%\begin{align}
%
%-\del_*^2 G = \delta(u_*-u_*')~.
%\label{eq:}
%
%\end{align}
%Such a Green's function is obtained from two homogeneous solutions. 
Then, the Green's function is given by
\begin{align}
G_V(u,u') 
= \left\{
\begin{array}{ll}
-\frac{1}{A} A_h(u)A_b(u') & (u'<u<1) 
\nonumber \\
-\frac{1}{A}A_h(u')A_b(u) & (0<u<u') 
\nonumber
\end{array}
\right.
%\label{eq:}
%
\end{align}
Thus,
\begin{align}
A_i^{(2)} 
&= a_i +  A_h\int_0^{u} du'\, A_b g_i(u') + A_b\int_{u}^{1} du'\, A_h g_i(u')~,
%\label{eq:}
%
\end{align}
and
\begin{subequations}
%\label{eq:}
\begin{align}
\del_u A_i^{(2)} 
&= \del_u a_i + \del_u A_h\int_0^{u} du'\, A_b g_i(u') +\del_u A_b\int_{u}^{1} du'\, A_h g_i(u')~, \\
2\del_u A_i^{(2)} |_{u=0} 
%&= 2\del_* \tilA_i^{(2)} |_{u=0} \\
&= 2\del_u a_i - 2\int_0^1 du'\, g_i(u')~.
%\label{eq:}
%
\end{align}
\end{subequations}
Then, the current is given by
\begin{subequations}
%\label{eq:}
\begin{align}
\bra\calJ_i^{(2)}\ket 
&= 2\del_u A_i^{(2)} -\frac{1}{2}q^2 A_i^{(2)}(\ln u-2\ln r_0)|_{u=0} \\
&= 2\del_u a_i - 2\int_0^1 du\, g_i(u') + \text{(counterterm)} \\
%&= 2\del_*a_i - 2\int_0^\infty du_*'\, g_i(u_*') + \text{(counterterm)} \\
&\sim \half q^2 \calA_i^{(2)} \{\ln u-(\ln u-2\ln r_0) \} - i\epsilon_i^{~j} q_j \int_0^1 \frac{du}{2u^2} |\Psi^{(1)}|^2 \\
&= q^2(\ln r_0)\calA_i^{(2)} - \frac{1}{4}i\epsilon_i^{~j} q_j |\psi^{(1)}|^2
%\int_0^1 \frac{du}{2u^2}\,\widetilde{|\Psi^{(1)}|^2}
\label{eq:current_meissner} \\
&=  \calJ_i^n +\calJ_i^s~.
%\label{eq:}
%
\end{align}
\end{subequations}
Here, we evaluate the integral using $U=F_0+\cdots$:
\begin{align}
\int_0^1 du\, \frac{U^2}{2u^2} = \frac{1}{4}~.
%\label{eq:}
%
\end{align}
The second term of \eq{current_meissner} is the supercurrent. 
%Once again, the supercurrent itself exists even under the Dirichlet boundary condition, but there is no Meissner effect. 
The first term of \eq{current_meissner} exists even for the pure Maxwell theory, and it is interpreted as the magnetization current due to the normal current. 

We impose the holographic semiclassical equation as the boundary condition:
\begin{subequations}
%\label{eq:}
\begin{align}
\del_j\calF^{ij} &=e^2\bra\calJ^i\ket~, \\
q^2 \calA_i^{(2)} &= e^2q^2 (\ln r_0) \calA_i^{(2)} + e^2\calJ_i^s \\
q^2(1-e^2 \ln r_0) \calA_i^{(2)} &= e^2 \calJ_i^s \\
q^2\calA_i^{(2)} &= \mu_m \calJ_i^s~, \\
\mu_m &= \frac{e^2}{1-e^2\ln r_0}~.
%\label{eq:}
%
\end{align}
\end{subequations}
$\B_2$ is then obtained as
\begin{align}
\B_2 = i\epsilon^{ij} q_i \calA_j^{(2)}
%&= -\mu_m \int_0^1 du\, \frac{\widetilde{|\Psi^{(1)}|^2}}{2u^2}\\
&= -\frac{1}{4}\mu_m|\psi^{(1)}|^2~.
%\label{eq:}
%
\end{align}
%The integral is evaluated as
%\begin{align}
%
%\int_0^1 du\, \frac{U^2}{2u^2} = \frac{1}{4}+\frac{1}{8}(B_0\ln2-2\epsmu) +O(H^2)~.
%\label{eq:}
%
%\end{align}
%where $A_t^{(0)}=(2+\epsmu)(1-u)$. 
Going back to the real space,
\begin{align}
\B_2
= c_1 - \frac{1}{4}\mu_m |\psi^{(1)}|^2~,
%\label{eq:}
%
\end{align}
where we add a zero mode solution $c_1$. 
The total $B$ is given by
\begin{align}
\B = \B_0 +\epsilon^2 \B_2
&= \Be - \frac{1}{4}\mu_m |\psi^{(1)}|^2~,
\label{eq:Hc2_2nd}
\end{align}
%(to the leading order in $B$) 
with $\Be:=\B_\infty$ and $\epsilon=1$.
Just like in the GL theory \eqref{eq:GL_meissner}, the magnetic induction $B$ reduces by the amount $|\psi^{(1)}|^2$, which implies the Meissner effect. The coefficient is consistent with $c_0=1/4$ determined in \sect{dual_GL}.

Under the Dirichlet boundary condition $e\to0$, $\mu_m=0$. Then, $\B=\Be$, so there is no Meissner effect. However, note that the supercurrent itself exists even under the Dirichlet boundary condition \eqref{eq:current_meissner}. 

%%------------------
%\subsubsection{Second order ($A_t$)}%\label{sec:}
%%------------------
\paragraph{The second order solution for $A_t^{(2)}$:}
To complete the second order analysis, solve the $A_t^{(2)}$ equation:
%The $A_t^{(2)}$ equation is given by
\begin{subequations}
%\label{eq:}
\begin{align}
0 &= \calL_t A_t^{(2)} - g_t~, \\
\calL_t &= \del_u^2 +\frac{1}{4uf} \del_i^2~, \\ 
g_t &= \frac{1}{4u^2f}J_t^{(2)}=\frac{1}{2u^2f}|\Psi^{(1)}|^2 A_t^{(0)}~.
%\label{eq:}
%
\end{align}
\end{subequations}
We impose the boundary conditions $A_t^{(2)}(u=0)=A_t^{(2)}(u=1)=0$.%
\footnote{One could impose the semiclassical equation as the boundary condition. But it is not necessary for $A_t$ here: The main reason why we impose the semiclassical equation on $A_i$ is to study the Meissner effect. } 
At $O(q^0)$, the solution is given by
\begin{align}
A_t^{(2)} 
& = \mu_c \frac{u(u-1)}{8(u+1)}  |\psi^{(1)}|^2  +O(q^2)~.
\label{eq:At2}
\end{align}
We utilize this solution below.

%%------------------
\subsection{Third order: the orthogonality condition and the free energy}%\label{sec:}
%%------------------

The construction so far has been discussed in Ref.~\cite{Natsuume:2022kic} in the context of the bulk 4-dimensional holographic superconductors. 
% v3.2
We now move to the third order. 
The third order is important because so far we solve the linear field equation for $\Psi$, so the normalization of $\Psi^{(1)}$ is not fixed. In other words, any configuration of vortex lattice is possible.

To fix the normalization, we take into account a nonlinear effect. The $O(\epsilon), O(\epsilon^3)$ equations are schematically written as
\begin{subequations}
%\label{eq:}
\begin{align}
\calL \Psi^{(1)} &= 0~,\\
\calL \Psi^{(3)} &= J^{(3)}~.
%\label{eq:}
%
\end{align}
\end{subequations}
Here,
\begin{subequations}
%\label{eq:}
\begin{align}
\calL &= D_{(0)}^2-m^2~,\\
J^{(3)} &= i\{D^M_{(0)}(A_M^{(2)}\Psi^{(1)})+A_M^{(2)}D^M_{(0)}\Psi^{(1)} \}~,
%\label{eq:}
%
\end{align}
\end{subequations}
where $D_M^{(0)}=\del_M-iA_M^{(0)}$. The $O(\epsilon), O(\epsilon^3)$ solutions satisfy \textit{the orthogonality condition}:
\begin{subequations}
\label{eq:orthogonality1}
\begin{align}
0 &= \int d^5x\sqrt{-g}\, \Psi^{(1) *} (\calL  \Psi^{(3)}-J^{(3)}) \\
&= \int d^5x\sqrt{-g}\,  \{ (\calL\Psi^{(1)})^* \Psi^{(3)}- \Psi^{(1) *}J^{(3)} \} \\
&= -\int d^5x\sqrt{-g}\, \Psi^{(1) *}J^{(3)} \\
&= \int d^5x\sqrt{-g}\, J_M^{(2)} A_{(2)}^M~.
%\label{eq:}
%
\end{align}
\end{subequations}
Recall
\begin{subequations}
%\label{eq:}
\begin{align}
J_i^{(2)} &= -\epsilon_i^{~j}\del_j|\Psi^{(1)}|^2~, \\
J_t^{(2)} &= -2|\Psi^{(1)}|^2 A_t^{(0)}~.
%\label{eq:}
%
\end{align}
\end{subequations}
Then, the orthogonality condition is rewritten as
\begin{align}
-2\int d\bmx du \sqrt{-g}\, g^{tt} |\Psi^{(1)}|^2 A_t^{(0)}A_t^{(2)}
= \int d\bmx du \sqrt{-g}\, g^{xx}|\Psi^{(1)}|^2 F_{xy}^{(2)}~.
\label{eq:orthogonality}
\end{align}

%The solutions $A_t^{(2)},A_y^{(2)}$ are formally written by Green's functions. We impose the boundary condition $A_t^{(2)}(u=1)=A_t^{(2)}(u=0)=0$. Then, $a_t=0$.
We evaluate this orthogonality condition. The left-hand side of \eq{orthogonality} is 
\begin{align}
(\text{LHS}) 
%= -2\int d\bmx du \sqrt{-g}\, g^{tt} |\Psi^{(1)}|^2 A_t^{(0)}A_t^{(2)}
%& = -2\int d\bmx du \sqrt{-g}\, g^{tt} |\Psi^{(1)}|^2 A_t^{(0)}\int_0^1 du' G_t(u,u')g_t(u') \\
& = - \frac{\mu_c^2}{192} \bra |\psi^{(1)}|^4 \ket +O(q^2)~,
%\label{eq:}
%
\end{align} 
Here, $\Psi^{(1)}= U(u)\psi^{(1)}(x,y)$ and we use \eq{At2} for $A_t^{(2)}$. $\bra \cdots \ket$ means the spatial integral.

For the right-hand side of \eq{orthogonality}, $A_y^{(2)}$ is obtained in \eq{Ay2}:
\begin{align}
F_{xy}^{(2)} &= \del_x A_y^{(2)} 
\to iq a_y-iq \int_0^1 du'\, G_V(u,u')g_y(u')~.
\label{eq:Fxy2}
\end{align}
The first term $a_y$ is the homogeneous solution obtained in \eq{ai}. At $O(q^0)$, $a_y = \calA_y^{(2)} + O(q^2)$, so $iq a_y = iq\calA_y^{(2)}=B_2$. 
%Using Green's functions, the left-hand side of the orthogonality condition \eqref{eq:orthogonality} is
The second term in \eq{Fxy2} is $O(q^2)$ because $g_y=O(q)$, so it can be ignored within our approximation: 
\begin{subequations}
%\label{eq:}
\begin{align}
(\text{RHS}) 
&= \int d\bmx du \sqrt{-g}\, g^{xx}|\Psi^{(1)}|^2 F_{xy}^{(2)}  \\
&= \int d\bmx du \sqrt{-g}\, g^{xx}|\Psi^{(1)}|^2 \left\{  iq a_y-iq \int_0^1 du'\, G_V(u,u')g_y(u') \right\}
\\
&= \int d\bmx\, iq \calA_y^{(2)} \int_0^1 du \sqrt{-g}\, g^{xx} |\Psi^{(1)}|^2 +O(q^2) 
%\nonumber \\
%&-iq \int d\bmx du \sqrt{-g}\, g^{xx}|\Psi^{(1)}|^2 \int_0^1 du'\, G_V(u,u')g_y(u')
\\
&=\frac{1}{4} \bra B_2|\psi^{(1)}|^2 \ket +O(q^2)~.
%\label{eq:}
%
\end{align}
\end{subequations}
Using the second-order result \eqref{eq:Hc2_2nd}, $B_2$ is given by
\begin{subequations}
%\label{eq:}
\begin{align}
\B &= \Bup+\B_2 =  \Be-\frac{1}{4}\mu_m |\psi^{(1)}|^2 \\
\to 
\B_2 &= \Be - \Bup -\frac{1}{4}\mu_m |\psi^{(1)}|^2
%\B &= \B_0+\B_2 = \Bup+\B_2 \\
%&= \Be-\frac{1}{4}\mu_m |\psi^{(1)}|^2 = \Be +\B_s~.
%\label{eq:}
%
\end{align}
\end{subequations}
%Then,
%\begin{align}
%
%\B_2 = \Be -\Bup+\B_s~.
%\label{eq:}
%
%\end{align}
Thus,
\begin{align}
\bra \B_2|\psi^{(1)}|^2 \ket
%&= \bra (\Be-\Bup+\B_s) |\psi^{(1)}|^2 \ket \\
& = (\Be-\Bup) \bra |\psi^{(1)}|^2 \ket -\frac{1}{4}\mu_m \bra |\psi^{(1)}|^4 \ket~.
%\label{eq:}
%
\end{align}

Then, the orthogonality condition \eqref{eq:orthogonality} becomes
\begin{align}
-\frac{\mu_c^2}{48} \bra |\psi^{(1)}|^4 \ket
& = (\Be-\Bup) \bra |\psi^{(1)}|^2 \ket -\frac{1}{4}\mu_m \bra |\psi^{(1)}|^4 \ket~.
%\label{eq:}
%
\end{align}
As discussed in \appen{vortex_GL}, the analogous relation in the GL theory is
\begin{align}
-\frac{b}{c_K}\bra |\psi^{(1)}|^4 \ket
& = (\Be-\Bup) \bra |\psi^{(1)}|^2 \ket - \mu_m c_K \bra |\psi^{(1)}|^4 \ket~.
%\label{eq:}
%
\end{align}
They agree because $\mu^2/48=1/12+O(\epsmu)$ and $b/c_K=1/12+O(\epsmu)$.

For the minimal holographic superconductor, the bulk scalar field has only the mass term. As is clear from the construction of the background solution, the nonlinearity comes from the backreaction of the bulk Maxwell field. The current analysis shows that the chemical potential $\mu_c$ actually plays the role of the nonlinear term $b$.

The rest of the analysis is the same as the GL theory. From the orthogonality condition, one gets
\begin{align}
\frac{b}{a}\frac{2\kappa^2-1}{2\kappa^2} \bra |\psi^{(1)}|^4 \ket = \left(1-\frac{\Be}{\Bup} \right) \bra |\psi^{(1)}|^2 \ket~,
%\label{eq:}
%
\end{align}
where we use
\begin{align}
\Bup= \frac{a}{c_K}~, \quad \kappa^2=\frac{b}{2\mu_m c_K^2}~.
%\label{eq:}
%
\end{align}
Introducing the Abrikosov parameter $\beta$, the orthogonality condition becomes
\begin{subequations}
%\label{eq:}
\begin{align}
\bra |\psi^{(1)}|^4 \ket &= \beta \bra |\psi^{(1)}|^2 \ket^2 \\
\to \frac{1}{2\kappa^2} \bra |\psi^{(1)}|^2 \ket &= \frac{a}{b}\frac{1-\frac{\Be}{\Bup}}{\beta(2\kappa^2-1)}~.
%\label{eq:}
%
\end{align}
\end{subequations}
For a Type II \SC, the vortex lattice is allowed when $\Be<\Bup$. In this case, $2\kappa^2-1$ must be positive. Namely, a Type II \SC\ is allowed when $\kappa^2>1/2$.

\paragraph{On-shell free energy:}
The on-shell action is given by
\begin{align}
\Sos + S_\text{bdy} = \int d^4 x\,  \frac{1}{2} \bra \calJ^{(2)}_i \ket \calA^{(2)}_i - \frac{1}{4e^2}\calF_{ij}^2~.
%\label{eq:}
%
\end{align}
The Maxwell part is
\begin{subequations}
%\label{eq:}
\begin{align}
%
%\calF^{(2)}_{xy} &= \epsilon^{ij}\del_i\calA^{(2)} _j \to \epsilon^{ij}iq_i\calA^{(2)}_j~,\\
(\calF^{(2)}_{ij})^2 &= 2q^2 (\calA^{(2)}_i)^2~.
%\label{eq:}
%
\end{align}
\end{subequations}
Combining with the normal current part
\begin{align}
 \calJ^{(2)}_i = q^2 \calA^{(2)}_i \ln r_0 +\calJ_i^s~,
%\label{eq:}
%
\end{align}
one obtains
\begin{align}
\Sos + S_\text{bdy} =\int d^4x\, \frac{1}{2}q^2 \left(\ln r_0-\frac{1}{e^2} \right) (\calA^{(2)}_i)^2 +\cdots
= \int d^4x\, -\frac{1}{4\mu_m} (\calF^{(2)}_{ij})^2 +\cdots~.
%\label{eq:}
%
\end{align}
Then, 
\begin{align}
-\beta F = \Sos + S_\text{bdy} 
= \int d^4 x\,  \frac{1}{2} \bra \calJ^s_i \ket \calA^{(2)}_i - \frac{1}{4\mu_m}\calF_{ij}^2~.
%\label{eq:}
%
\end{align}
The GL on-shell free energy is also written in this form [see \eq{fOS_Hc2}]. As a result,
\begin{itemize}
\item 
The rest of the analysis is the same as the GL theory.
\item
The favorable vortex lattice configuration is the one with the minimum $\beta$. As is well-known in the GL theory, the minimum is $\beta\simeq 1.16$ given by the triangular lattice.
\end{itemize}

%%%%%%%%%
\section{Nonminimal holographic superconductors}\label{sec:nonminimal}
%%%%%%%%%

Ref.~\cite{Herzog:2010vz} studies the analytic solution for a class of nonminimal holographic superconductors (St\"{u}ckelberg holographic superconductor) based on suggestions of Refs.~\cite{Franco:2009if,Franco:2009yz}:
\begin{subequations}
%\label{eq:}
\begin{align}
S_\text{m} &= -\frac{1}{g^2} \int d^5x \sqrt{-g} \biggl\{ \frac{1}{4}F_{MN}^2 + K|D_M\Psi|^2+V \biggr\}~,\\
K &= 1+A|\Psi|^2~,\quad
V= m^2 |\Psi|^2+B |\Psi|^4~.
%\label{eq:}
%
\end{align}
\end{subequations}
% that
There are 2 features that one can realize:
\begin{enumerate}
\item
The arbitrary values may not be allowed for $A$ and $B$. If $A<0$, $\Psi$ may become a (unitary) ghost. If $B<0$, the potential may not be bounded below. For simplicity, we set $A,B>0$.
\item
The new terms appear as nonlinear terms for $\Psi$. 
For the minimal holographic superconductor, the nonlinearity comes from the backreaction of the bulk Maxwell field on $\Psi$ at $O(\epsilon^3)$. Thus, one expects that $A$ and $B$ affect the analysis at $O(\epsilon^3)$. 
In the dual GL theory, the nonlinearity comes from the $|\psi|^4$-potential. Thus, one expects that $A$ and $B$ affect $b$, the coefficient of $|\psi|^4$.
%Then, these terms do not affect the dual GL theory at the linear order and affect the GL theory at nonlinear orders. 
We will see this explicitly.
\end{enumerate}
The bulk equations of motion are given by
\begin{subequations}
%\label{eq:}
\begin{align}
0 &= -D^M(KD_M\Psi) + \frac{\del V}{\del \Psi^*} + \frac{\del K}{\del \Psi^*} |D_M\Psi|^2~, \\
0 &= \nabla_NF^{MN} - J^M~,\\
J_M &= -iK\{ \Psi^* D_M\Psi -\Psi(D_M\Psi)^*\}~.
%= 2\Im(\Psi^* D_M\Psi)~.
%\label{eq:}
%
\end{align}
\end{subequations}
The goal of this section is to identify the dual GL theory for this class of nonminimal holographic superconductors. 

%%------------------
\subsection{High-temperature phase}%\label{sec:low}
%%------------------

%%------------------
%\subsubsection{The order parameter response function}%\label{sec:}
%%------------------
\paragraph{The order parameter response function (high temperature):}
We consider the linear perturbation $\delta\Psi$ of the form  $e^{iqx}$. 
At high temperatures, $\delta A_t$ and $\delta A_i$ decouple from the $\delta\Psi$-equation, and it is enough to consider the $\delta\Psi$-equation:
%The bulk scalar perturbation equation is given by
\begin{align}
0=\del_u\left( \frac{f}{u}\del_u \delta\Psi \right) + \left[ \frac{A_t^2}{4u^2 f} -\frac{q^2}{4u^2} + \frac{1}{u^3} \right] \delta\Psi~,
\label{eq:scalar_high_non}
\end{align}
where $A_t =(2+\epsmu)(1-u)$. The field equation remains the same as the minimal case. 
%In the high-temperature phase, $\epsmu<0$. Set $\epsmu\to l \epsmu, q^2 \to l q^2$, and expand $\delta\Psi$ as a series in $l$:
%\begin{align}
%
%\delta\Psi = F_0+l F_1+l^2F_2+\cdots~.
%\label{eq:}
%
%\end{align}
Then, one obtains the response function $\chi_>$, the correlation length $\xi_>$, and the thermodynamic susceptibility $\chi_>^T$:
\begin{subequations}
%\label{eq:response_high}
\begin{align}
J &=  \frac{q^2-2\epsmu}{4}\Cone~, \\
\to \chi_> &=\frac{\del \delta\psi}{\del J} = \frac{4}{q^2-2\epsmu} \propto \frac{1}{q^2+\xi_>^{-2}}~, \\
\xi_>^2  &=-q^{-2} = \frac{1}{-2\epsmu}~, \\
\chi_>^T &= \left.\frac{\del\delta\psi}{\del J}\right|_{q=0} =\frac{2}{-\epsmu}~.
%\label{eq:}
%
\end{align}
\end{subequations}

%%------------------
%\subsubsection{The upper critical magnetic field $\Bup$}%\label{sec:}
%%------------------
\paragraph{The upper critical magnetic field $\Bup$:}
We apply a magnetic field $\B$ and approach the critical point from the high-temperature phase. 
Near $\Bup$, $\Psi$ remains small, and one can expand matter fields as a series in $\epsilon$:
\begin{subequations}
%\label{eq:}
\begin{align}
\Psi(\vecx,u) &= \epsilon\Psi^{(1)}+\cdots~, \\
A_t(\vecx,u) &= A_t^{(0)}+\epsilon^2 A_t^{(2)}+\cdots~, \\
A_y(\vecx,u) &= A_y^{(0)}+\epsilon^2 A_y^{(2)}+\cdots~.
%\label{eq:}
%
\end{align}
\end{subequations}
At zeroth order,
\begin{align}
A_t^{(0)} = \mu(1-u)~,\quad
A_x^{(0)} = 0~,\quad
A_y^{(0)} = \B x~.
%\label{eq:}
%
\end{align}
At first order, the bulk scalar equation for $\Psi^{(1)}$ remains the same as the minimal case. Using the ansatz $\Psi^{(1)}=\chi(x)\rhou(u)$, the solution for $\chi$ is given by Hermite function, and the $U$-equation takes 
the same form as \eq{scalar_high_non} with the replacement $\B\to q^2$, so we immediately conclude
\begin{align}
\Bup = \frac{1}{-\xi_>^2}~.
%\label{eq:}
%
\end{align}
We consider the \HSC\ with scalar mass $m^2=-4$, but \textit{the above relation holds exactly for this class of nonminimal \HSCs\ with arbitrary mass.}
%to all orders in the perturbative expansion in $\epsmu$.} 
Thus,
\begin{align}
\Bup &= 2\epsmu+\cdots~.
%2(1-\ln2)\epsmu^2+\cdots~.
%\label{eq:}
%
\end{align}

%%------------------
\subsection{Low-temperature phase}%\label{sec:low}
%%------------------

%%------------------
%\subsubsection{The background}\label{sec:background_nonminimal}
%%------------------
\paragraph{The background:}
%\margin{checked}
One can construct the low-temperature background as in the minimal case \cite{Herzog:2010vz}:
\begin{subequations}
%\label{eq:}
\begin{align}
A_t(u) &= A_t^{(0)}+\epsilon^2 A_t^{(2)}+\epsilon^4 A_t^{(4)} +\cdots~, \\
\Psi(u) &= \epsilon\Psi^{(1)}+ \epsilon^3\Psi^{(3)} +\cdots~.
%\label{eq:}
%
\end{align}
\end{subequations}
The background solution remains the same as the minimal case up to $O(\epsilon^2)$:
\begin{subequations}
%\label{eq:}
\begin{align}
A_t^{(0)} &=\mu_c(1-u)~, \\
\Psi^{(1)} &= -\frac{u}{1+u}~, \\%\quad\text{at}\quad \mu_c=\Delta=2~, \\
A_t^{(2)} &= \mu_2(1-u) - \frac{u(1-u)}{4(1+u)}~,
%&\sim \mu_2 + \frac{1}{4}(-1-4\mu_2) u +\cdots~,
%\quad(u\to0)
%\label{eq:}
%
\end{align}
\end{subequations}
where $\mu_2$ is an integration constant, but it is fixed at the next order from the source-free condition of $\Psi^{(3)}$:
\begin{subequations}
%\label{eq:}
\begin{align}
\Psi^{(3)} &= \frac{(1-2A+B)u^2}{12(1+u)^2} 
+\frac{1+4A+4B-24\mu_2}{96} \frac{u\ln u}{1+u} 
+\frac{8\mu_2-1+4A}{16}\frac{u\ln(1+u)}{1+u} \\
&\sim \frac{1+4A+4B-24\mu_2}{96} u\ln u + \cdots~,
\quad(u\to0)~.
%\label{eq:}
%
\end{align}
\end{subequations}
The source of the order parameter is given by
\begin{align}
J^{(3)} = \frac{1+4A+4B-24\mu_2}{48}~.
%\label{eq:}
%
\end{align}
Then, for the spontaneous condensate $J=0$, 
\begin{align}
\mu_2= \frac{1+4A+4B}{24}~.
\label{eq:mu2_non}
\end{align}
This fixes the overall constant $\epsilon$ of the condensate:
\begin{subequations}
%\label{eq:}
\begin{align}
\mu &= 2+ \epsilon^2 \mu_2+\cdots~, \\
\epsilon_\mu &:=\mu-2 = \mu_2\epsilon^2 +\cdots~, \\
\epsilon^2 &=\frac{1}{\mu_2}\epsmu+\cdots = \frac{24}{1+4A+4B}\epsmu+\cdots~.
%\label{eq:}
%
\end{align}
\end{subequations}
%The construction so far has been discussed in \cite{Herzog:2010vz,Natsuume:2018yrg}. 
The higher order expressions are too cumbersome to write here, and we only give the asymptotic forms. At $O(\epsilon^4)$,
\begin{align}
A_t^{(4)} &\sim \mu_4(1-u) + \left\{ \frac{5-37A-10B}{288}+\frac{-1+4A+8\mu_2}{32}\ln 2 \right\} u +\cdots~.
\label{eq:mu4_non}
\end{align}
Again, $\mu_4$ is an integration constant, but it is fixed at the next order.

%%------------------
%\subsubsection{The on-shell free energy}%\label{sec:}
%%------------------
\paragraph{The on-shell free energy:}
The on-shell free energy has been discussed in Ref.~\cite{Herzog:2010vz}. 
For the nonminimal case, the on-shell matter action is given by
\begin{align}
\Sos= -\int d^4x\, \calA_t A_t^{(+)} + \int d^5x\, \sqrt{-g} g^{tt}g^{uu} A_t^2 |\Psi|^2 
+  \int d^5x\, \sqrt{-g}\left[ B|\Psi|^4+A|\Psi|^2|D_M\Psi|^2 \right]~.
%\label{eq:}
%
\end{align}
For the $\Psi\neq 0$ solution, the on-shell action becomes
\begin{align}
\frac{\Sos_{\Psi\neq0}}{\beta V_3} = 4(1+\mu_2)\epsilon^2+ \epsilon^4\left( 4\mu_4+ \mu_2^2-\frac{\mu_2}{4}+\frac{1+4A+4B}{48}\right) +\cdots.
%\label{eq:}
%
\end{align}
For the $\Psi= 0$ solution,
the on-shell action becomes
\begin{align}
\frac{\Sos_{\Psi=0}}{\beta V_3} &= \mu^2 = 4(1+\mu_2)\epsilon^2+ \epsilon^4\left( 4\mu_4+ \mu_2^2 \right) +\cdots.
%\label{eq:}
%
\end{align}
Thus, the difference of the on-shell action is given by
\begin{subequations}
%\label{eq:}
\begin{align}
\delta \Sos &= \Sos_{\Psi\neq0}-\Sos_{\Psi=0} \\
&= \frac{1+4A+4B-12\mu_2}{48}\epsilon^4 \times \beta V_3+\cdots \\
&= \frac{1+4A+4B}{96}\epsilon^4 \times \beta V_3+\cdots \\
&= -\delta f_\psi \times\beta V_3~, \\
\delta f_\psi &= -\frac{1+4A+4B}{96}\epsilon^4 = -\frac{6}{1+4A+4B}\epsmu^2~.
%\label{eq:}
%
\end{align}
\end{subequations}

%%------------------
%\subsubsection{The critical magnetic field $\Bc$ and the GL parameter}%\label{sec:}
%%------------------
\paragraph{The critical magnetic field $\Bc$ and the GL parameter:}
The bulk Maxwell action does not change from the minimal case, so the on-shell free energy when $\B\neq0$ remains the same as \eq{SosH}:
\begin{subequations}
%\label{eq:}
\begin{align}
\delta f_B &= -\frac{1}{2\mu_m}\B^2~, \\% \times \beta V_3~,\\
\delta f_\psi &= -\frac{6}{1+4A+4B}\epsmu^2~. %\times \beta V_3~.
%\label{eq:}
%
\end{align}
\end{subequations}
Then, the critical magnetic field $\Bc$ is given by
\begin{align}
\Bc^2 =\frac{12}{1+4A+4B}\mu_m\epsmu^2~.
%\label{eq:}
%
\end{align}
The GL parameter is then given by
\begin{align}
\kappa_B^2 &:= \frac{1}{2}\left(\frac{\Bup}{\Bc}\right)^2 = \frac{1+4A+4B}{6\mu_m}~.
%\label{eq:}
%
\end{align}

%%------------------
%\subsubsection{The penetration length}%\label{sec:}
%%------------------
\paragraph{The penetration length:}
Consider the perturbation of the form $A_y \propto e^{iqx}$. The bulk Maxwell equation becomes
\begin{align}
0=\del_u(f\del_u A_y) -\left( \frac{q^2}{4u} + K\frac{|\Psi|^2}{2u^2} \right) A_y~.
%\label{eq:}
%
\end{align}
Again we solve the integral equation iteratively and obtains
\begin{subequations}
%\label{eq:}
\begin{align}
\bra \calJ^y \ket &= -2\calA_y\int_0^1 du\, \left( \frac{q^2}{4u}+ K\frac{|\Psi|^2}{2u^2} \right) +\cdots +(\text{counterterm}) \\
&= \left\{ q^2(\ln r_0) - \frac{1}{2}\epsilon^2 +\cdots \right\} \calA_y  \\
&=: (c_n q^2 -c_s\epsilon^2) \calA_y~.
%\label{eq:}
%
\end{align}
\end{subequations}
However, $K$ makes no contribution at $O(\epsilon^2)$. Then, $\lambda$ remains the same as the minimal case when expressed in terms of $\epsilon$:
\begin{subequations}
%\label{eq:}
\begin{align}
\lambda^2 &= \frac{1}{\mu_m c_s\epsilon^2} = \frac{2}{\mu_m \epsilon^2} =\frac{1+4A+4B}{12\mu_m\epsmu}~, \\
\mu_m &= \frac{e^2}{1-c_ne^2}~.
%\label{eq:}
%
\end{align}
\end{subequations}
%The GL parameter is then given by
%\begin{align}
%
%\kappa^2 = \frac{\lambda^2}{-\xi_>^2} = \frac{1+4A+4B}{6\mu_m}~.
%\label{eq:}
%
%\end{align}

%%------------------
%\subsubsection{The order parameter response function}%\label{sec:}
%%------------------
\paragraph{The order parameter response function (low temperature):}
We take the gauge $A_u=0$ and perturb around the low-temperature background:
\begin{subequations}
%\label{eq:}
\begin{align}
\Psi &= \bmPsi+\delta\Psi~, \\
A_t &= \bmA_t+a_t~, \\
A_x &= 0 +a_x~,
%\label{eq:}
%
\end{align}
\end{subequations}
where boldface letters indicate the background. We consider the perturbation of the form $e^{iqx}$. 
The $\delta\Psi$ equation is real, so $\delta\Psi^*=\delta\Psi$. Then, one can set $a_x=0$.
Set $\epsilon\to l \epsilon, q \to l q$, %(\omega\to l^2 \omega)
 and expand the fields as a series in $l$:
\begin{subequations}
%\label{eq:}
\begin{align}
a_t &= a_t^{(0)} + l a_t^{(1)} + l^2 a_t^{(2)}+\cdots~, \\
\delta\Psi &= F_0 + l F_1+ l^2F_2 +\cdots~.
%a_x &= (1-u^2)^{-i\omega/4} (a_x^{(0)}+l a_x^{(1)} + l^2 a_x^{(2)}+\cdots)~.
%\label{eq:}
%
\end{align}
\end{subequations}
Here, $a_t|_{u=0}=\delta\at$. Below we give the $\delta\at=0$ solution for simplicity. The $A,B$-dependences appear only in the $F_2$ equations. Thus, the solution remains the same as the minimal case except $F_2$:
\begin{align}
F_0 &= -\Cone\,\frac{u}{1+u}~, \quad a_t^{(0)}=a_t^{(2)}=0~, \quad a_t^{(1)} = - \Cone\,\epsilon \frac{u(1-u)}{2(1+u)}~.
%\label{eq:}
%
\end{align}
Up to $O(l^2)$, the asymptotic form of the solution is given by
\begin{subequations}
%\label{eq:}
\begin{align}
a_t &\sim -\half \Cone\,\epsilon u~, \\
\delta\Psi &\sim  \frac{6q^2+\epsilon^2(1+4A+4B)}{48}\Cone\,  u\ln u- \Cone\, u~, \\
&\sim  \frac{1}{8} \Cone\,(q^2+4\epsmu)u\ln u -\Cone\, u~.
%\label{eq:}
%
\end{align}
\end{subequations}
Then, the order parameter response function remains the same as the minimal case when expressed in terms of $\epsmu$:
\begin{subequations}
%\label{eq:}
\begin{align}
J &=  \frac{q^2+4\epsmu}{4}\Cone~, \\
\to \chi_< &=\frac{\del \delta\psi}{\del J} = \frac{4}{q^2+4\epsmu} \propto \frac{1}{q^2+\xi_<^{-2}}~, \\
\xi_<^2  &=-q^{-2} = \frac{1}{4\epsmu}~, \\
\chi_<^T &= \left.\frac{\del\delta\psi}{\del J}\right|_{q=0} =\frac{1}{\epsmu}~.
%\label{eq:}
%
\end{align}
\end{subequations}
%The charge density is given by
%\begin{align}
%
%\bra \delta\calJ^t \ket 
%&= - 2a_t' = \Cone\epsilon~.
%\label{eq:}
%
%\end{align}

%%------------------
%\subsubsection{The vortex lattice}%\label{sec:}
%%------------------
\paragraph{The vortex lattice:}
As in the minimal case in \sect{vortex}, we expand matter fields as a series in $\epsilon$:
\begin{subequations}
%\label{eq:}
\begin{align}
\Psi(\vecx,u) &= \epsilon\Psi^{(1)}+ \epsilon^3\Psi^{(3)}+\cdots~, \\
A_t(\vecx,u) &= A_t^{(0)}+\epsilon^2 A_t^{(2)}+\cdots~, \\
A_i(\vecx,u) &= A_i^{(0)}+\epsilon^2 A_i^{(2)}+\cdots~.
%\label{eq:}
%
\end{align}
\end{subequations}
Even for the nonminimal case, the analysis remains the same up to $O(\epsilon^2)$. The difference arises at $O(\epsilon^3)$. 

At $O(\epsilon)$, the normalization of $\Psi^{(1)}$ is not fixed, and one needs to take into account a nonlinear effect. The $O(\epsilon), O(\epsilon^3)$ equations are given by
\begin{subequations}
%\label{eq:}
\begin{align}
\calL \Psi^{(1)} &= 0~,\\
\calL \Psi^{(3)} &= J^{(3)}~.
%\label{eq:}
%
\end{align}
\end{subequations}
 The $O(\epsilon), O(\epsilon^3)$ solutions satisfy the orthogonality condition:
%Again, in order to fix the normalization of $\Psi^{(1)}$, we consider the $O(\epsilon^3)$ equation. We evaluate the orthogonality condition \eqref{eq:orthogonality1}:
\begin{align}
0 &= -\int d^5x\sqrt{-g}\, \Psi^{(1) *}J^{(3)}~.
%\label{eq:}
%
\end{align}
For the minimal holographic superconductor, the orthogonality condition gives
\begin{align}
-\frac{\mu_c^2}{48} \bra |\psi^{(1)}|^4 \ket
& = (\B-\Bup) \bra |\psi^{(1)}|^2 \ket -\frac{1}{4}\mu_m \bra |\psi^{(1)}|^4 \ket~.
%\label{eq:}
%
\end{align}
For the nonminimal case, the left-hand side of the above equation is replaced by
\begin{align}
-\frac{24A+16B+(1-2A)\mu_c^2}{48} \bra |\psi^{(1)}|^4 \ket
= -\frac{1+4A+4B}{12} \bra |\psi^{(1)}|^4 \ket~.
%\label{eq:}
%
\end{align}

\begin{itemize}
\item
In the GL theory, this left-hand side of the analogous relation \eqref{eq:orthogonality3} has the coefficients $-b/c_K=-(1+4A+4B)/12$, so the bulk analysis agrees with the dual GL theory. 
\item
For the minimal case, the nonlinearity comes from the backreaction of the bulk Maxwell field. As a result, $\mu_c$ plays the role of the nonlinear term $b$. For the nonminimal case, the parameters $A,B$ as well as $\mu_c$ play the role of the nonlinear term $b$.
\item
The rest of the analysis remains the same, so the triangular lattice is the most favorable configuration.
\end{itemize}

%%------------------
\subsection{The dual GL theory}%\label{sec:}
%%------------------

Following the minimal holographic superconductor analysis, one obtains
\begin{align}
b_0 = \frac{\bo}{48}~,
%\label{eq:}
%
\end{align}
and  the dual GL theory is given by
\begin{align}
f= \frac{1}{4}|D_i\psi|^2-\frac{\epsilon_\mu}{2}|\psi|^2+\frac{\bo}{96}|\psi|^4+\frac{1}{4\mu_m}\calF_{ij}^2-(\psi J^*+\psi^* J)~.
% \right\}~.
%\label{eq:}
%
\end{align}
From the GL theory, one can obtain physical quantities and they all agree with the bulk results:
\begin{subequations}
%\label{eq:}
\begin{align}
|\psi_0|^2 &= \frac{24}{\bo}\epsmu~, \decrease \\
\delta f_\text{OS} &= -\frac{6}{\bo}\epsmu^2~, \decrease \\
\Bup &= 2\epsmu~, \\
\Bc^2 &= \frac{12}{\bo}\mu_m\epsmu^2~, \decrease \\
\xi_<^2 &= \frac{1}{-2\epsmu}~, \\
\xi_>^2 &= \frac{1}{4\epsmu}~, \\
\lambda^2 &= \frac{\bo}{12\mu_m \epsmu}~, \increase \\
\kappa^2 &= \frac{\bo}{6\mu_m}~, \increase \\
A_> &=2A_<~.
%\label{eq:}
%
\end{align}
\end{subequations}
Here, the arrows indicate the behaviors when $A,B>0$ (at a fixed chemical potential $\epsmu$.)
%Let us fix the chemical potential $\epsmu$.
One can understand the $A,B$-dependences as follows: 
\begin{enumerate}
\item The net effect of $A,B$ is to make $b$ larger (the coefficient of the $|\psi|^4$ term). 
\item Then, the condensate $\epsilon$ becomes smaller.
\item The penetration length $\lambda$ is the same as the minimal case when expressed by $\epsilon$, but $\epsilon$ becomes smaller which makes $\lambda$ larger for a fixed $\epsmu$.
\item The correlation lengths do not change, but $\lambda$ becomes larger, which makes the GL parameter $\kappa^2$ larger. Namely, the system approaches a more Type II \SC-like material.
\item This implies that $\Bc^2$ becomes smaller since $\Bup^2$ remains the same. 
\item In this analysis, only the combination $A+B$ appears in the dual GL theory, but there is no reason to expect that only this combination appears in general away from the critical point. 
\end{enumerate}

%%%%%%%%%
\section{Discussion}%\label{sec:}
%%%%%%%%%

In this paper, we analyze a class of holographic superconductors. We compute various physical quantities in the bulk theory, and they all agree with the GL theory. In this way, we identify the dual GL theory analytically. 
\begin{itemize}
\item
The relation $\Bup=1/(-\xi_>^2)$ is well-known in the GL theory, but we find that the relation holds \textit{exactly} for the \HSCs\ 
% that
that we consider.
\item
However, we are not claiming that the relation is exact for real \SCs. Rather, this may come from the strong coupling limit. In the strong coupling limit, we learned that one often encounters universal relations using the holographic duality. Here, the universality does not mean the universality classes found in field theories. Some examples are
\begin{itemize}
\item
$\eta/s=1/(4\pi)$, where $\eta$ is the shear viscosity and $s$ is the entropy density \cite{Kovtun:2004de}.
\item
% v2
The holographic chaos and pole-skippings: the Green's functions are not uniquely determined at pole-skipping points in the complex momentum space, and the locations of pole-skipping points are always located at Matsubara frequencies (see, \eg, Refs.~\cite{Grozdanov:2017ajz,Blake:2018leo,Grozdanov:2019uhi,Blake:2019otz,Natsuume:2019xcy}).
%v2
The pole-skipping was originally discussed in the context of holographic chaos \cite{Shenker:2013pqa,Roberts:2014isa,Roberts:2014ifa,Shenker:2014cwa,Maldacena:2015waa}.
\end{itemize}
The relation may be another example of the universality. 

\item
Our results correspond to the strong coupling limit, so it would be interesting to take into account finite-coupling corrections and to see how the relation and various parameters change under the corrections \cite{second}. 

The holographic duality has two couplings, 't~Hooft coupling $\lambda$ and the number of colors $N_c$. Our results correspond to the large-$N_c$ limit, \ie, $\lambda\to\infty, N_c\to\infty$. In the bulk theory, the $1/\lambda$-corrections correspond to higher-derivative corrections or $\alpha'$-corrections. The $1/N_c$-corrections correspond to string loop corrections or quantum gravity corrections. 
\item
In this paper, we focus on a class of  holographic superconductors. 
% v3.2
But there exist other analytic solutions \cite{Herzog:2009ci,Herzog:2010vz,Natsuume:2018yrg}, and it would be interesting to carry out a similar analysis for the solutions.  
\item
Also, it is interesting to carry out numerical computations and to see how the results deviates from analytic results as the system is away from the critical point. 
\item
We take the probe limit $g^2N_c^2 \gg 1$. It is interesting to take the backreaction into account to see how our analytic results change. It is difficult to study the system analytically, so one would need a numerical analysis.
\end{itemize}

%%%%%%%%%
\section*{Acknowledgments}%\label{sec:}
%%%%%%%%%

%%%%%%%%%
%\acknowledgments

I would like to thank Takashi Okamura for his continuous suggestions and interest throughout the work.
I also would like to thank \'Oscar Dias, and Gary Horowitz for useful discussions.
This research was supported in part by a Grant-in-Aid for Scientific Research (17K05427) from the Ministry of Education, Culture, Sports, Science and Technology, Japan. 

%\newpage
\appendix

\small

%%%%%%%%%
\section{Restoring dimensions}\label{sec:dimensions}
%%%%%%%%%

% v3.2
In the text, we set $r_0=L=g=1$ for simplicity, but we restore the dimensions in this appendix.
In a scale-invariant theory, the only scale is the temperature $T$, so one expects that $r_0$ and $L$ appear in the form $T\sim r_0/L^2$ in the boundary physical quantities, but let us check this explicitly. 

The bulk action is given by
\begin{subequations}
%\label{eq:}
\begin{align}
S_\text{bulk} &= \frac{1}{16\pi G_5}\int d^5x \sqrt{-g}(R-2\Lambda)+S_\text{m}~, \\
S_\text{m} &= -\frac{1}{g^2L} \int d^5x \sqrt{-g} \biggl\{ \frac{1}{4}F_{MN}^2 + L^2(|D_M\Psi|^2+m^2 |\Psi|^2) \biggr\}~.
%\label{eq:}
%
\end{align}
\end{subequations}
Here, we choose the mass dimensions as $[A_M]=\mass$, $[\Psi]=\mass^2$, and $[g]=\mass^0$. 

\paragraph{Dictionary:}
In the coordinate $\tilu=(L/r)^2$, the metric is given by
\begin{subequations}
%\label{eq:}
\begin{align}
ds_5^2 &= \left( \frac{r}{L}\right)^2(-fdt^2+dx^2+dy^2+dz^2)+\frac{dr^2}{r^2f} \\
&=\frac{1}{\tilu}(-fdt^2+dx^2+dy^2+dz^2)+L^2\frac{d\tilu^2}{4\tilu^2f}~.
%f &= 1-\left(\frac{r_0}{r}\right)^4=1-u^2~, 
%\label{eq:}
%
\end{align}
\end{subequations}
The asymptotic behaviors of matter fields are given by
\begin{subequations}
%\label{eq:}
\begin{align}
A_\mu & \sim \tilcalA_\mu  + \tilA_\mu^{(+)} \tilu~, \\
\Psi &\sim \frac{1}{2}\tilPsi^{(-)} \tilu \ln \tilu +\tilPsi^{(+)} \tilu~.
\end{align}
\end{subequations}
Using the standard procedure, one obtains
\begin{subequations}
%\label{eq:}
\begin{align}
\bra J^t \ket &= -\frac{2}{g^2L^2} \tilA_t^{(+)}+(\text{counterterm})~, \\
\bra J^i \ket &= \frac{2}{g^2L^2} \tilA_i^{(+)}+(\text{counterterm})~, \\
J &=\tilPsi^{(-)}~,\\
\psi =\bra \calO \ket &= -\frac{1}{g^2}\tilPsi^{(+)}~.
%\label{eq:}
%
\end{align}
\end{subequations}
The coordinate $u=(r_0/r)^2$ is related to $\tilu$ by
\begin{align}
u= \left( \frac{r_0}{L}\right)^2\tilu~.
%\label{eq:}
%
\end{align}
Then, in the $u$-coordinate, \eg,
\begin{subequations}
%\label{eq:}
\begin{align}
A_\mu & \sim \tilcalA_\mu  + \tilA_\mu^{(+)} \tilu 
= \tilcalA_\mu  + \tilA_\mu^{(+)} \left( \frac{L}{r_0}\right)^2 u \\
&=:  \calA_\mu + A_\mu^{(+)} u~,
\end{align}
\end{subequations}
so that 
\begin{subequations}
%\label{eq:}
\begin{align}
\bra J^t \ket &= -\frac{2}{g^2L^2} \left( \frac{r_0}{L}\right)^2 A_t^{(+)}+(\text{counterterm})~, \\
\bra J^i \ket &= \frac{2}{g^2L^2} \left( \frac{r_0}{L}\right)^2 A_i^{(+)}+(\text{counterterm})~, \\
J &= \left( \frac{r_0}{L}\right)^2 \Psi^{(-)}~,\\
\psi &= \bra \calO \ket = -\frac{1}{g^2}\left( \frac{r_0}{L}\right)^2 \Psi^{(+)}~.
\label{eq:dict_scalar}
\end{align}
\end{subequations}

The counterterm is given by
\begin{align}
S_\text{CT}= -\int d^4x\, \frac{1}{4g^2}\sqrt{-\gamma}\gamma^{\mu\nu}\gamma^{\rho\sigma}F_{\mu\rho}F_{\nu\sigma} \times \ln\tilu^{1/2}~,
%\label{eq:CT}
%
\end{align}
where $\gamma_{\mu\nu}$ is the 4-dimensional boundary metric. The log term is rewritten as
\begin{align}
\ln \tilu^{1/2} =\ln u^{1/2}-\ln \left(\frac{r_0}{L}\right)~.
%\label{eq:}
%
\end{align}

\paragraph{Dimensions:}
One can restore $r_0$ and $L$ from the scaling analysis and the dimensional analysis.
The pure AdS geometry is invariant under the scaling
\begin{align}
x^\mu \to a x^\mu~, \quad
\tilu \to a^2 \tilu~.
%\label{eq:}
%
\end{align}
This gives  the scaling dimensions as
\begin{align}
[x]_s=-1, \quad
[\tilu]_s=-2, \quad
[r_0]_s=1, \quad
[L]_s=0~.
%\label{eq:}
%
\end{align}
On the other hand, the mass dimensions are
\begin{align}
[x]=\mass^{-1}~, \quad
 [\tilu]=\mass^0~, \quad
 [r_0]=\mass^{-1}~, \quad
 [L]=\mass^{-1}~.
%\label{eq:}
%
\end{align}
Note that the scaling dimensions and the mass dimensions differ for $\tilu,r_0,$ and $L$.  
The temperature has the following dimensions:
\begin{align}
T \sim \frac{r_0}{L^2} \to [T]_s=1, [T]=\mass~.
%\label{eq:}
%
\end{align}

From the bulk point of view, the scaling is just a coordinate transformation. The bulk Maxwell field is a one-form, and $\Psi$ is a scalar, so they transform as
\begin{align}
A_\mu \to A_\mu/a~,\quad
 \Psi \to \Psi~.
%\label{eq:}
%
\end{align}
Namely, the scaling dimensions are $[A_\mu]_s=1$ and $[\Psi]_s=0$. 
Then, one obtains
\begin{subequations}
%\label{eq:}
\begin{align}
&[\calA_\mu]_s = 1~, [J^\mu]_s = 3~, \\
&[J]_s = [\psi]_s = 2~.
%\label{eq:}
%
\end{align}
\end{subequations}
The mass dimensions are 
\begin{subequations}
%\label{eq:}
\begin{align}
&[\calA_\mu] = \mass~, [J^\mu] = \mass^3~, \\
&[J] = [\psi] = \mass^2~.
%\label{eq:}
%
\end{align}
\end{subequations}
Namely, the mass dimensions coincide with the scaling dimensions.

Then, for example
\begin{itemize}
\item The critical point $\mu_c=2$ has the scaling dimension 1 and the mass dimension 1, so
\begin{align}
\mu_c=2 \to 
 \mu_c = 2 \left( \frac{r_0}{L^2}\right)~.
%\label{eq:}
%
\end{align}

\item
The condensate $\psi$ has the scaling dimension 2 and the mass dimension 2, so
\begin{align}
\psi \sim  \epsmu^{1/2} 
\to \psi \sim \left( \frac{r_0}{L^2}\right)^{3/2} \epsmu^{1/2}~.
%\label{eq:}
%
\end{align}

\item
The correlation length has the scaling dimension $-1$ and the mass dimension $-1$, so
\begin{align}
\xi^2 \sim \frac{1}{\epsmu}
\to \xi^2 \sim \frac{L^2}{r_0} \frac{1}{\epsmu}~.
%\label{eq:}
%
\end{align}
A similar result applies to the penetration length $\lambda$. However, one has the UV divergence and needs the holographic renormalization for $\lambda$, so the scaling is broken by the $\ln(r_0/L)$ term.

\end{itemize}

\paragraph{Bulk equations:}
Let us restore dimensions explicitly. In the $u$-coordinate, the metric is given by
\begin{align}
ds_5^2 %&= r^2(-fdt^2+dx^2+dy^2+dz^2)+\frac{dr^2}{r^2f} \\
&=\left( \frac{r_0}{L}\right)^2\frac{1}{u}(-fdt^2+dx^2+dy^2+dz^2)+L^2\frac{du^2}{4u^2f}~.
%f &= 1-\left(\frac{r_0}{r}\right)^4=1-u^2~, 
%\label{eq:}
%
\end{align}
The field equations are given by
\begin{subequations}
%\label{eq:}
\begin{align}
0 &= \del_u\left( \frac{f}{u}\del_u \Psi \right) + \left[ \frac{L^4}{r_0^2} \frac{A_t^2}{4u^2f} + \frac{L^4}{r_0^2} \frac{1}{4u^2} (\del_i-iA_i)^2 - \frac{m^2L^2}{4u^3} \right] \Psi~, \\
0 &=\del_u^2 A_t - L^4 \frac{|\Psi|^2}{2u^2f}A_t +  \frac{L^4}{r_0^2} \frac{1}{4uf}\del_i^2A_t~, \\
0 &=\del_u(f \del_u A_y)- L^4 \frac{|\Psi|^2}{2u^2}A_y +  \frac{L^4}{r_0^2} \frac{1}{4u}\del_i^2A_y~.
%\label{eq:}
%
\end{align}
\end{subequations}
In the bulk equations, $r_0$ and $L$ appear in the combination
\begin{align}
\bA_M = \frac{L^2}{r_0} A_M~, \quad
\bq = \frac{L^2}{r_0} q~, \quad
\bPsi = L^2 \Psi~.
%\label{eq:}
%
\end{align}
The``$\bar{~~}$" variables are all dimensionless (the scaling dimensions and the mass dimensions). 
In the``$\bar{~~}$" variables, the bulk equations reduce to the ones with $r_0=L=1$. Then, all our results in the text are valid in the ``$\bar{~~}$" variables. In the ``$\bar{~~}$" variables , the AdS/CFT dictionary becomes
\newcommand{\bJ}{\bar{J}}
\newcommand{\bcalJ}{\bar{\calJ}}
\begin{subequations}
%\label{eq:}
\begin{align}
\bPsi &\sim \frac{\bJ}{2} u\ln u-\bpsi u~,\\
\bJ &=\biggl( \frac{L^2}{r_0} \biggr)^2 J~, \quad
\bpsi =\biggl( \frac{L^2}{r_0} \biggr)^2 \psi~,\\
%\label{eq:}
%
%\end{align}
%\end{subequations}
%and 
%\begin{subequations}
%\label{eq:}
%\begin{align}
%
\bA_i &\sim \bcalA_i + \frac{\bcalJ^i}{2} u~,\\
\bcalA_i &=\biggl( \frac{L^2}{r_0} \biggr) \calA_i~,  \quad
\bcalJ^i =\biggl( \frac{L^2}{r_0} \biggr)^3 \calJ^i~.
%\label{eq:}
%
\end{align}
\end{subequations}
For example, 
\begin{itemize}
\item
The critical point is given by
\begin{align}
\bmu_c = \bA_t|_{u=0} = 2 
\to \mu_c = 2 \left( \frac{r_0}{L^2}\right)~.
 %\label{eq:}
%
\end{align}

\item
The condensate is given by
\begin{align}
\bPsi^{(+)} %= \epsilon 
\sim \bepsmu^{1/2}
%\to \psi \sim \frac{1}{g^2L^2} \left( \frac{r_0}{L}\right)^2\bepsmu^{1/2}~.
\to \psi \sim \frac{1}{g^2} \left( \frac{r_0}{L^2}\right)^{3/2} \epsmu^{1/2}~.
%\label{eq:}
%
\end{align}
%where we use \eq{dict_scalar}.

\item
The correlation length at high temperature is given by
\begin{subequations}
%\label{eq:}
\begin{align}
-\bq^2 &=-2\bepsmu \\%-2(1-\ln2)\bepsmu^2 +\cdots \\
\to \xi^2 &= -\frac{1}{q^2}= -\left( \frac{L^2}{r_0} \right)^2 \frac{1}{\bq^2} = \left( \frac{L^2}{r_0} \right)^2 \frac{1}{-2\bepsmu}
 = \left( \frac{L^2}{r_0} \right)  \frac{1}{-2\epsmu}~. 
%\label{eq:}
%
\end{align}
\end{subequations}

\end{itemize}

\paragraph{The dual GL theory:}
%We set $\psi:= g^2 \bra\psi\ket$. 
The bulk results are written by dimensionless quantities, so the dual GL theory should be written by dimensionless quantities as well:
\begin{align}
\bar{f} = c_K |\bD_i\bpsi|^2-a|\bpsi|^2+ \frac{b}{2} |\bpsi|^4 %+c_6|\bpsi|^6 +\cdots 
+\frac{1}{4\mu_m}\bcalF_{ij}^2+\cdots~.
%\label{eq:}
%
\end{align}
%where we include only the $|\bpsi|^6$ as a higher order term as an example. 
Here,
\begin{subequations}
%\label{eq:}
\begin{align}
\bar{f} &= \left( \frac{L^2}{r_0}\right)^4 f~,\quad
\bar{x} = \frac{r_0}{L^2} x~,\quad
\bcalA_i =  \frac{L^2}{r_0} \calA_i~, \quad
\bpsi =  \left( \frac{L^2}{r_0}\right)^2 \psi~, \\
\bD_i &= \bdel_i - i\bcalA_i~, \quad
\bcalF_{ij}= \bdel_i\bcalA_j - \bdel_j\bcalA_i~, \quad
a= a_0\bepsmu(1+\cdots)~.
%\label{eq:}
%
\end{align}
\end{subequations}
For example, $|\bpsi|^2 \sim \bepsmu$ which is consistent with the bulk result. 
In terms of the variables without ```$\bar{~~}$",
\begin{align}
f= c_K \left( \frac{L^2}{r_0}\right)^2 |D_i\psi|^2 - a|\psi|^2
+ \frac{b}{2} \left( \frac{L^2}{r_0}\right)^4 |\psi|^4 
%+c_6 \left( \frac{L^2}{r_0}\right)^8 |\psi|^6 +\cdots 
+\frac{1}{4\mu_m}\calF_{ij}^2+\cdots~.
%\label{eq:}
%
\end{align}
Finally, redefine $\psi$ as
\begin{align}
|\phi|^2 =  c_K \left(\frac{L^2}{r_0}\right)^2 |\psi|^2
%\label{eq:}
%
\end{align}
so that $\phi$ has the canonical mass dimension 1 and the canonical normalization:
\begin{align}
f &= |D_i \phi|^2 - \tila \left( \frac{r_0}{L^2}\right)^2|\phi|^2
+ \frac{\tilb}{2} |\phi|^4 
%+c_6 \left( \frac{L^2}{r_0}\right)^2 |\cpsi|^6 +\cdots 
+\frac{1}{4\mu_m}\calF_{ij}^2+\cdots~,
%\label{eq:}
%
\end{align}
where $\tila=a/c_K$ and $\tilb=b/c_K^2$.

%%%%%%%%%
\section{Supplementary information of the vortex lattice}%\label{sec:vortex_GL}
%%%%%%%%%

%%------------------
\subsection{GL analysis}\label{sec:vortex_GL}
%%------------------

Here, we summarize the conventional GL analysis near $\Bup$ for the reader's convenience.
The field equations are given by
\begin{subequations}
%\label{eq:}
\begin{align}
0 &= -c_K D^2\psi-a\psi+b\psi|\psi|^2~, \\
0 &=\del_j\calF^{ij} - \mu_m\calJ^i~, \\
\calJ_i &= - ic_K [\psi^* D_i\psi-\psi (D_i\psi)^* ]~.
%\label{eq:}
%
\end{align}
\end{subequations}

Near the upper critical magnetic field $\Bup$, $\psi$ remains small, and one can expand matter fields as a power series:
\begin{subequations}
%\label{eq:}
\begin{align}
\psi &= \epsilon \psi^{(1)} + \cdots~, \\
\calA_i &= \calA_i^{(0)}+\epsilon^2 \calA_i^{(2)} + \cdots~.
%\label{eq:}
%
\end{align}
\end{subequations}
At zeroth order, the Maxwell equation is 
$0= \del_j \calF_{(0)}^{ij}~,$
%\begin{align}
%
%0= \del_j \calF_{(0)}^{ij}~,
%\label{eq:}
%
%\end{align}
so one has a homogeneous magnetic field
$\calA_y^{(0)}=B_0x.$
%\begin{align}
%
%\calA_y^{(0)}=B_0x~.
%\label{eq:}
%
%\end{align}
At first order, the order parameter field obeys
\begin{align}
0= -c_K(\del_i-i \calA_i^{(0)})^2\psi^{(1)} - a \psi^{(1)}~.
%\label{eq:}
%
\end{align}
Using the ansatz
$\psi^{(1)} = e^{iqy}\chi_q(x),$
%\begin{align}
%
%\psi^{(1)} = e^{iqy}\chi_q(x)~,
%\label{eq:}
%
%\end{align}
the first-order equation becomes
\begin{align}
c_K\left\{ -\del_x^2+B_0^2 \left(x-\frac{q}{B_0}\right)^2 \right\} \chi_q  = a\chi_q~.
%\label{eq:}
%
\end{align}
This is the Landau problem, and the solution is given by the Hermite function $H_n$ as
\begin{align}
\chi_q = e^{-z^2/2}H_n(z)~, \quad
z:= \sqrt{B_0}\left(x-\frac{q}{B_0}\right)~.
%\label{eq:}
%
\end{align}
The eigenvalue is given by
\begin{align}
E_n=(2n+1)B_0 = \frac{a}{c_K}~.
%\label{eq:}
%
\end{align}
$B_0$ takes the maximum value when $n=0$. This $B_0$ gives the upper critical magnetic field $\Bup=B_0(n=0)=a/c_K$. 

The general solution is written as
\begin{align}
\psi^{(1)} &= \int_{-\infty}^\infty dq\, C(q) e^{iqy} \chi_q(x)~.
\label{eq:psi_0}
\end{align}
%Because we solve the linear equation, one cannot fix the overall normalization at this moment. We have to take into account the nonlinear effect.

The first order solution \eqref{eq:psi_0} satisfies
\begin{align}
(\del_y - i \calA_y^{(0)}) \psi^{(1)} = i(\del_x - i \calA_x^{(0)}) \psi^{(1)}~,
%\label{eq:}
%
\end{align}
so
\begin{subequations}
%\label{eq:}
\begin{align}
\calJ^{(2)}_x &= 2c_K\Im \left[(\psi^{(1)})^* D_x^{(0)}\psi^{(1)} \right] = -c_K\del_y|\psi^{(1)}|^2~,\\
\calJ^{(2)}_y &=  c_K\del_x|\psi^{(1)}|^2~,
%\label{eq:}
%
\end{align}
\end{subequations}
or 
\begin{align}
\calJ^{(2)}_a = 2c_K\Im \left[(\psi^{(1)})^* D_a^{(0)}\psi^{(1)} \right] = -c_K\epsilon_a^{~b} \del_b |\psi^{(1)}|^2~,
\label{eq:GL_current}
\end{align}
where the Latin indices $a,b$ run though $x$ and $y$, and $\epsilon_{xy}=1$. 
Then, at second order, one can integrate the equation:
\begin{subequations}
%\label{eq:}
\begin{align}
0 &= \del^b \calF_{ab}^{(2)} - \mu_m \calJ_a^{(2)} \\
&= \epsilon_{ab} \del^b(\calF_{xy}^{(2)}+ c_K\mu_m |\psi^{(1)}|^2)~, \\
\calF_{xy}^{(2)} &= c_1-\mu_m c_K |\psi^{(1)}|^2~.
%\label{eq:}
%
\end{align}
\end{subequations}
Asymptotically, $|\psi^{(1)}|\to0$, so $\calF_{xy}=B\to B_0+c_1=\Be$. Then,
\begin{align}
\calF_{xy} &= B = \Be -\mu_m c_K |\psi^{(1)}|^2~.
\label{eq:GL_meissner}
\end{align}
Thus, the magnetic induction $\B$ reduces by the amount $|\psi^{(1)}|^2$ which implies the Meissner effect. %We discuss its holographic counterpart in \sect{upper}.% in the text.

So far we solve the linear field equation for $\psi$, so the normalization of $\psi^{(1)}$ is not fixed. To fix the normalization, we take into account a nonlinear effect. 
The $O(\epsilon), O(\epsilon^3)$ equations are given by
\begin{subequations}
%\label{eq:}
\begin{align}
0 &= \calL \psi^{(1)}~, \\
0 &= \calL \psi^{(3)} - J^{(3)}~, \\
\calL &= -c_K (D_i^{(0)})^2-a~. 
%J^{(3)} &= -(2A_i^{(0)}A_i^{(2)}+b|\psi^{(1)}|^2)\psi^{(1)}~.
%\label{eq:}
%
\end{align}
\end{subequations}
The $O(\epsilon), O(\epsilon^3)$ solutions satisfy the orthogonality condition:
\begin{subequations}
%\label{eq:}
\begin{align}
0 &= \int d^2x\, \psi^{(1)*}(\calL\psi^{(3)}-J^{(3)}) \\
&= \int d^2x\, (\calL\psi^{(1)})^* \psi^{(3)}- \psi^{(1)*}J^{(3)} \\
&= \int d^2x\, - \psi^{(1)*}J^{(3)}~.
%\label{eq:}
%
\end{align}
\end{subequations}
Here,
\begin{align}
- \psi^{(1)*}J^{(3)} &= -c_K \psi^{(1) *}(D_i^{(0)}D_i^{(2)}+D_i^{(2)}D_i^{(0)}) \psi^{(1)} + b |\psi^{(1)}|^4~.
%\label{eq:}
%
\end{align}
The first term is written as
\begin{subequations}
%\label{eq:}
\begin{align}
&ic_K \psi^{(1) *}(D_i^{(0)}\calA_i^{(2)}+\calA_i^{(2)}D_i^{(0)}) \psi^{(1)} \\
&\to -ic_K \{ -(D_i^{(0)}\psi^{(1)})^*\psi^{(1)}+\psi^{(1) *}D_i^{(0)}\psi^{(1)}\} \calA_i^{(2)} \\
& = - \calJ_i^{(2)} \calA_i^{(2)}
%\label{eq:}
%
\end{align}
\end{subequations}
so that 
\begin{align}
b\bra |\psi^{(1)}|^4 \ket= \bra \calJ_i^{(2)} \calA_i^{(2)} \ket~,
\label{eq:orthogonality2}
\end{align}
where $\bra \cdots \ket$ means the spatial integral.

$B_2$ is written as
\begin{subequations}
%\label{eq:}
\begin{align}
B&= \Bup+B_2 = \Be-\mu_m c_K |\psi^{(1)}|^2 \\
\to B_2 &= \Be-\Bup-\mu_m c_K |\psi^{(1)}|^2
%B&= \Be-\mu_m c_K |\psi^{(1)}|^2 = \Be+B_s \\
%&= B_0+B_2 = \Bup+B_2
%\label{eq:}
%
\end{align}
\end{subequations}
%Then,
%\begin{align}
%
%B_2 = \Be -\Bup+B_s~.
%\label{eq:}
%
%\end{align}
Recall $ \calJ_i^{(2)}= -c_K\epsilon_i^{~j} \del_j |\psi^{(1)}|^2$. Then, 
\begin{subequations}
\label{eq:orthogonality3}
\begin{align}
b\bra |\psi^{(1)}|^4 \ket
&= \bra \calJ_i^{(2)}\calA_i^{(2)} \ket \\
&= -c_K \bra B_2 |\psi^{(1)}|^2 \ket \\
%&= -c_K \bra(\Be-\Bup+B_s) |\psi^{(1)}|^2 \ket \\
& =-c_K (\Be-\Bup) \bra |\psi^{(1)}|^2 \ket + \mu_m c_K^2 \bra |\psi^{(1)}|^4 \ket~.
%\label{eq:}
%
\end{align}
\end{subequations}
One then obtains
\begin{align}
\frac{b}{a}\frac{2\kappa^2-1}{2\kappa^2} \bra |\psi^{(1)}|^4 \ket = \left(1-\frac{\Be}{\Bup} \right) \bra |\psi^{(1)}|^2 \ket~,
%\label{eq:}
%
\end{align}
where we use
\begin{align}
\Bup= \frac{a}{c_K}, \kappa^2=\frac{b}{2\mu_m c_K^2}~.
%\label{eq:}
%
\end{align}
Introducing the Abrikosov parameter $\beta$ as
\begin{subequations}
%\label{eq:}
\begin{align}
\bra |\psi^{(1)}|^4 \ket &= \beta \bra |\psi^{(1)}|^2 \ket^2 \\
\to \frac{1}{2\kappa^2} \bra |\psi^{(1)}|^2 \ket &= \frac{a}{b}\frac{1-\frac{\Be}{\Bup}}{\beta(2\kappa^2-1)}~.
%\label{eq:}
%
\end{align}
\end{subequations}
For a Type II \SC, the vortex lattice is allowed when $\Be<\Bup$. In this case, $2\kappa^2-1$ must be positive. Namely, a Type II \SC\ is allowed when $\kappa^2>1/2$.

The on-shell free energy is given by
\begin{align}
f_\text{OS} = -\frac{1}{2}b \bra |\psi^{(1)}|^4 \ket + \frac{1}{4\mu_m} \calF_{ij}^2 = -\frac{1}{2} \bra \calJ^{(2)}_i \calA^{(2)}_i \ket + \frac{1}{4\mu_m} \calF_{ij}^2 
\label{eq:fOS_Hc2}
\end{align}
using the orthogonality condition, and this agrees with the bulk result.

%%------------------
\subsection{Summary of the bulk analysis}\label{sec:vortex_formula}
%%------------------

The analysis of the vortex lattice is rather involved, so we collect the necessary formulae 
% that
that one needs to evaluate. 
We slightly generalize the argument under the following assumptions:
\begin{enumerate}
\item
We consider the minimal \HSC\ in a SAdS$_5$-like background.
\item
But we do not use the explicit form of $f(u)$. 
\item 
We assume that the bulk Maxwell equations take the same form as the SAdS$_5$ case.
\end{enumerate}
Then, the vortex lattice analysis reduces to evaluate several integrals.

We expand
\begin{subequations}
%\label{eq:}
\begin{align}
\Psi(\vecx,u) &= \epsilon\Psi^{(1)}+ \epsilon^3\Psi^{(3)}+\cdots~, \\
A_t(\vecx,u) &= A_t^{(0)}+\epsilon^2 A_t^{(2)}+\cdots~, \\
A_i(\vecx,u) &= A_i^{(0)}+\epsilon^2 A_i^{(2)}+\cdots~.
%\label{eq:}
%
\end{align}
\end{subequations}
At the zeroth order, 
\begin{align}
A_t^{(0)} = \mu_c(1-u)~, 
A_x^{(0)} = 0~, 
A_y^{(0)} = B_0x =\Bup x~.
%\label{eq:}
%
\end{align}
For the first order solution, one can use separation of variables:
\begin{align}
\Psi^{(1)}=U(u)\psi^{(1)}(x,y)~.
%\label{eq:}
%
\end{align}

\paragraph{The second order solution for $A_i^{(2)}$:}
The Maxwell equation at second order is given by
\begin{subequations}
%\label{eq:}
\begin{align}
0 &= \calL_V A_i^{(2)} - g_i~,\\
\calL_V &= \del_u(f\del_u)-\frac{q^2}{4u}~, \\
g_i &= i\epsilon_i^{~j} q_j \frac{|\Psi^{(1)}|^2}{4u^2}~.
%\label{eq:}
%
\end{align}
\end{subequations}
Using the bulk Green's function $G_V$, the solution is formally written as
\begin{align}
A_i^{(2)} &= a_i - \int_0^1 du'\, G_V(u,u')g_i(u')~.
%\label{eq:Ay2}
%
\end{align}
The first term $a_i$ is the homogeneous solution.
Obtain 2 independent homogeneous solutions $A_b, A_h$ at $O(q^0)$. The solution $A_b$ satisfies the boundary condition at the AdS boundary and $A_h$ satisfies the boundary condition at the horizon. 
\begin{align}
W &:= A_b\del_uA_h-(\del_uA_b)A_h =: \frac{A}{f}~.
%\label{eq:}
%
\end{align}
Then,
\begin{subequations}
%\label{eq:}
\begin{align}
\del_u A_i^{(2)} 
&= \del_u a_i +  \frac{\del_u A_h}{A}\int_0^{u} du'\, A_b g_i(u') +\frac{\del_u A_b}{A}\int_{u}^{1} du'\, A_h g_i(u')~, \\
2\del_u A_i^{(2)} |_{u=0} 
&= 2\del_u a_i +2\frac{\del_u A_b}{A}\int_0^1 du'\, A_h(u')g_i(u')~, \\
2\del_u a_i &= \frac{q^2}{2}\calA_i^{(2)}\frac{\ln u}{f}+\cdots
%\label{eq:}
%
\end{align}
\end{subequations}
If the current is given by the standard AdS/CFT dictionary, the supercurrent becomes
\begin{subequations}
%\label{eq:}
\begin{align}
\bra\calJ_i \ket 
&=2\del_u A_i^{(2)} +(\text{counterterm}) |_{u=0} = \calJ_i^{s} +\calJ_i^{n}~, \\
\calJ_i^{s} 
&=- i\epsilon_i^{~j} q_j |\psi^{(1)}|^2 \times I_1~, \\
I_1 &=: - \frac{\del_u A_b(0)}{A} \int_0^1 du'\,\frac{A_h U^2}{2u'^2}~.
%\label{eq:}
%
\end{align}
\end{subequations}

The homogeneous solution represents the normal current but needs the holographic renormalization. The counterterm is
\begin{align}
(\text{CT}) &= -\del_j(\sqrt{-\gamma}F_{(2)}^{ij}) \times \half c_T (\ln u-2\ln r_0) \\
&= -q^2 \calA_i^{(2)} \left( \frac{-g_{tt}}{g_{xx}} \right)^{1/2} \times  \half c_T (\ln u-2\ln r_0)~,
%&= -q^2 \calA_i^{(2)} f^{1/2} \times  \half c_T (\ln u-2\ln r_0)~,
%\label{eq:}
%
\end{align}
where we use the gauge $\del_i \calA^i=0$. Then, the normal current is
\begin{subequations}
%\label{eq:}
\begin{align}
\bra\calJ_i^{n}\ket 
%&= \left. \half q^2 \calA_i^{(2)} \left[ \frac{\ln u}{f} (1 - c_T f ^{3/2}) +2c_T f^{1/2}\ln r_0 \right] \right|_{u=0} \\
&= \left. \half q^2 \calA_i^{(2)} \left[ \ln u \left\{ \frac{1}{f} - c_T \left( \frac{-g_{tt}}{g_{xx}} \right)^{1/2} \right\} 
+ 2c_T \left( \frac{-g_{tt}}{g_{xx}} \right)^{1/2} \ln r_0 \right] \right|_{u=0} \\
&= \frac{1}{f(0)} (\ln r_0) q^2 \calA_i^{(2)} =: c_n q^2 \calA_i^{(2)}~, \\
c_n &= \frac{1}{f(0)}\ln r_0~,\\
c_T &= \left. \frac{1}{f}\left( \frac{-g_{tt}}{g_{xx}} \right)^{-1/2} \right|_{u=0}~.
%\label{eq:}
%
\end{align}
\end{subequations}
%Here we choose $c_T=1/f(0)^{3/2}$ in order to cancel the UV divergence. 

The holographic semiclassical equation then gives
\begin{subequations}
%\label{eq:}
\begin{align}
\del_j\calF^{ij} &=e^2\bra\calJ^i\ket~, \\
q^2 \calA_i^{(2)} &= e^2q^2 c_n \calA_i^{(2)} + e^2\calJ_i^s \\
q^2 \calA_i^{(2)} &=  \mu_m \calJ_i^s~, \\
\mu_m &= \frac{e^2}{1-e^2c_n}~.
%\label{eq:}
%
\end{align}
\end{subequations}
$ B_2 = i\epsilon^{ij} q_i \calA_j^{(2)}$, and the total $B$ is given by
\begin{align}
B = B_0 +\epsilon^2 B_2
&= \Be - \mu_m I_1  |\psi^{(1)}|^2~.
%\label{eq:Hc2_2nd}
%
\end{align}
The above relation should reduce to the analogous relation in the GL theory:
\begin{align}
B = B_0 +\epsilon^2 B_2
&= \Be - \mu_m c_0  |\psi^{(1)}|^2~.
%\label{eq:Hc2_2nd}
%
\end{align}
Namely, the magnetic induction $B$ reduces by the amount $|\psi^{(1)}|^2$, which implies the Meissner effect. 

\paragraph{The second order solution for $A_t^{(2)}$:}
Similarly, solve the $A_t^{(2)}$ equation:
\begin{subequations}
%\label{eq:}
\begin{align}
0 &= \calL_t A_t^{(2)} - g_t~, \\
\calL_t &= \del_u^2 +\frac{1}{4uf} \del_i^2~, \\ 
g_t &= \frac{1}{2u^2f}|\Psi^{(1)}|^2 A_t^{(0)}~.
%\label{eq:}
%
\end{align}
\end{subequations}
%where $A_t^{(0)}=\mu_c(1-u)$.
The solution is formally written as 
\begin{align}
A_t^{(2)} &=  C_1(1-u) - \int_0^1 du'\, G_t(u,u')g_t(u')~.
%a_t&=C_1(1-u)
%\label{eq:}
%
\end{align}
Two independent homogeneous solutions at $O(q^0)$ are
\begin{subequations}
%\label{eq:}
\begin{align}
A_b &=u~, \quad
A_h =1-u~, \\
W &:= A_b\del_u A_h-(\del_u A_b)A_h=-1=A~.
%\label{eq:}
%
\end{align}
\end{subequations}
Then, 
\begin{subequations}
%\label{eq:}
\begin{align}
A_t^{(2)} 
& = C_1(1-u) -A_h\int_0^{u} du'\, A_b g_t(u') - A_b\int_{u}^{1} du'\, A_h g_t(u') \\
& = (1-u)\int_0^{1} du'\, (1-u') g_t(u') -(1-u)\int_0^{u} du'\, g_t(u') - \int_{u}^{1} du'\, (1-u') g_t(u')~,\\
& =: \mu_c |\psi^{(1)}|^2 \times I_t~.
%\label{eq:}
%
\end{align}
\end{subequations}

\paragraph{Third order:}
The orthogonality condition is given by
\begin{subequations}
%\label{eq:}
\begin{align}
-2\mu_c^2 \bra |\psi^{(1)}|^4 \ket \times I_L
&= \bra B_2 |\psi^{(1)}|^2 \ket \times I_R~, \\
I_L &= \int_0^1du\, \sqrt{-g} g^{tt}U^2(1-u)I_t~, \\
I_R &= \int_0^1du\, \sqrt{-g} g^{xx}U^2~.
%\label{eq:}
%
\end{align}
\end{subequations}
Using the $B_2$ result
\begin{subequations}
%\label{eq:}
\begin{align}
\B &= \Bup+\B_2 =  \Be-\mu_m c_0 |\psi^{(1)}|^2 \\
\to 
\B_2 &= \Be - \Bup -\mu_m c_0 |\psi^{(1)}|^2~,
%\label{eq:}
%
\end{align}
\end{subequations}
the orthogonality condition becomes
\begin{align}
-2\mu_c^2 \frac{I_L}{I_R} \bra |\psi^{(1)}|^4 \ket
& = (\Be-\Bup) \bra |\psi^{(1)}|^2 \ket -\mu_m c_0 \bra |\psi^{(1)}|^4 \ket~. %\\
%I_L &= \int_0^1du\, \sqrt{-g} g^{xx}U^2~, \\
%I_R &= \int_0^1du\, \sqrt{-g} g^{tt}U^2(1-u)I_t~,
%\label{eq:}
%
\end{align}
The above orthogonality condition should reduce to the analogous relation in the GL theory:
\begin{align}
-\frac{b_0}{c_0}\bra |\psi^{(1)}|^4 \ket
& = (\Be-\Bup) \bra |\psi^{(1)}|^2 \ket - \mu_m c_0 \bra |\psi^{(1)}|^4 \ket~.
%\label{eq:}
%
\end{align}
The rest of the analysis is the same as the GL theory, and the favorable vortex lattice configuration is the triangular lattice.

To summarize, what one needs to evaluate are 4 integrals:
\begin{align}
I_1, I_t,I_L,I_R~.
%\label{eq:}
%
\end{align}

\end{document}